\documentclass[preprint]{imsart}

\usepackage{amsthm,amsmath,natbib}
\RequirePackage[colorlinks,citecolor=blue,urlcolor=blue]{hyperref}
\usepackage{amsmath,amsthm,amsbsy}
\usepackage{amsfonts}
\usepackage{amssymb}
\usepackage{graphicx}
\usepackage{epsfig}
\usepackage{subfigure}
\usepackage{float}
\usepackage{multirow}

\newcommand{\R}{\mathbb{R}}

\newcommand {\argmin} {\mathop{\rm argmin}}

\arxiv{arXiv:0000.0000}

\startlocaldefs
\numberwithin{equation}{section}
\theoremstyle{plain}

\endlocaldefs

\begin{document}

\begin{frontmatter}
\title{Principal nested shape space analysis of molecular dynamics data\thanksref{T1}}
\runtitle{Principal nested shape spaces}

\begin{aug}
\author{\fnms{Ian L.} \snm{Dryden}
\ead[label=e1]{ian.dryden@nottingham.ac.uk}},
\author{\fnms{Kwang-Rae} \snm{Kim}
\ead[label=e2]{kr4kim@gmail.com}},
\author{\fnms{Charles A.} \snm{Laughton}
\ead[label=e3]{charles.laughton@nottingham.ac.uk}}
\&
\author{\fnms{Huiling} \snm{Le}
\ead[label=e4]{huiling.le@nottingham.ac.uk}}

\thankstext{T1}{This work was supported by the Engineering and Physical Sciences Research Council [grant number EP/K022547/1] and Royal Society Wolfson Research Merit Award WM110140.}

\runauthor{I.L. Dryden et al.}

\affiliation{University of Nottingham\thanksmark{m1} and SAS Korea\thanksmark{m2}}

\address{School of Mathematical Sciences,\\
University of Nottingham,\\
University Park,\\
Nottingham, NG7 2RD,\\
United Kingdom\\
\printead{e1}\\
\phantom{E-mail:\ }\printead*{e4}}

\address{SAS Korea\\
South Korea\\
\printead{e2}}

\address{School of Pharmacy,\\
University of Nottingham,\\
University Park,\\
Nottingham, NG7 2RD,\\
United Kingdom\\
\printead{e3}}

\end{aug}

\begin{abstract}
Molecular dynamics simulations produce huge datasets of temporal 
sequences of 
molecules. It is of interest to summarize the shape 
evolution of the molecules in a 
succinct, low-dimensional 
representation. However, Euclidean techniques such as 
principal components analysis (PCA) can be problematic as the data may 
lie far from in a flat manifold. Principal nested spheres gives a fundamentally different 
decomposition of data from the usual Euclidean sub-space based PCA 
\citep{JDM12}. Sub-spaces of successively lower dimension are
 fitted to the data in a backwards manner, with the aim of retaining signal and 
dispensing with noise at each stage. We adapt the methodology to 3D sub-shape spaces 
and provide some practical fitting algorithms. The methodology is applied to cluster 
analysis 
of peptides, where different states of the molecules can be identified. Also, the temporal transitions 
between cluster states are explored. 
\end{abstract}


\end{frontmatter}

\section{Introduction}
There are many notions of shape, and one of the most common is that the shape 
of an object is obtained by removing location, rotation and scale \citep{Kendall84}. Analyzing the shapes of objects 
measured at sets of labelled landmarks is of interest in a wide variety of disciplines \citep{Drydmard16}, and in many applications the dataset is large, either in sample size or in dimension, or in both. For example, 
in molecular dynamics simulations the three dimensional co-ordinates of 
a molecule of hundreds or thousands of atoms may be available for millions of observations. It is of interest to
summarize the main features of shape variability, such as describing the main modes of variability 
and clustering the data into several states. A key aspect is to project the data into a low dimensional 
space which retains the main features of the signal in the data. 

The most common approaches to dimension reduction in shape analysis involve projecting the 
data into a tangent space to the mean shape and then carrying out principal components analysis (PCA) in this Euclidean 
space \citep[Section 7.7]{Kent94,Cootetal94,Drydmard16}. This approach has been successful in many
applications over the past couple of decades, although if the data are very dispersed then 
the method can be problematic as the data may lie far from in a flat manifold. In addition, linear variation
(corresponding to geodesics in the manifold) may not be the most appropriate or efficient summary of the 
variability. There have been several advances in dimension reduction on manifolds, motivated by shape
analysis applications. \cite{FLPJ04} developed principal geodesic analysis to the manifold setting
by finding principal directions and variances in the tangent space and projecting back to the manifold
using the exponential map.  The above approaches are forwards fitting, in that the lower dimension representations
are fitted before the higher dimensional ones. 

 \cite{HZ06} proposed a method of PCA for Riemannian manifolds
based on geodesics of the intrinsic metric, and
\cite{HHM10} developed an algorithmic method to perform intrinsic PCA on quotient spaces based on generalized geodesics.
\cite{KDL10} proposed two intrinsic methods of fitting minimal geodesics through shape data, and an extension to 
polynomial fitting. These intrinsic methods differ in their formulation from the tangent space methods, in that
tangent space PCA involves maximising the explained variance for each linear component in the tangent space, whereas the 
intrinsic geodesic methods involves minimizing the unexplained variance in order to fit a geodesic subspace. The intrinsic
methods are also forwards fitting in the sense that the first principal geodesic is fitted before the second principal geodesic etc., although 
there is a different notion of a mean  which is the point that minimizes the variance of the projected data in the principal geodesic.
\cite{Panaretosetal14} introduce principal flow which is a curve passing through 
the mean of the data, where a particle moves along the principal flow in a path of maximal variation, up to smoothness constraints. 

A very different backwards approach to subspace fitting was considered by \cite{JDM12} who 
introduced a technique called Principal Nested Spheres (PNS)
for sequentially decomposing a unit sphere into subspheres of successively lower dimension. This backwards 
approach to dimension reduction is the approach that we consider, which requires special adaptation 
to be applicable to three dimensional shapes and to large datasets, and these are our main methodological contributions. 

\begin{figure}[htbp]
\includegraphics[scale = 0.4]{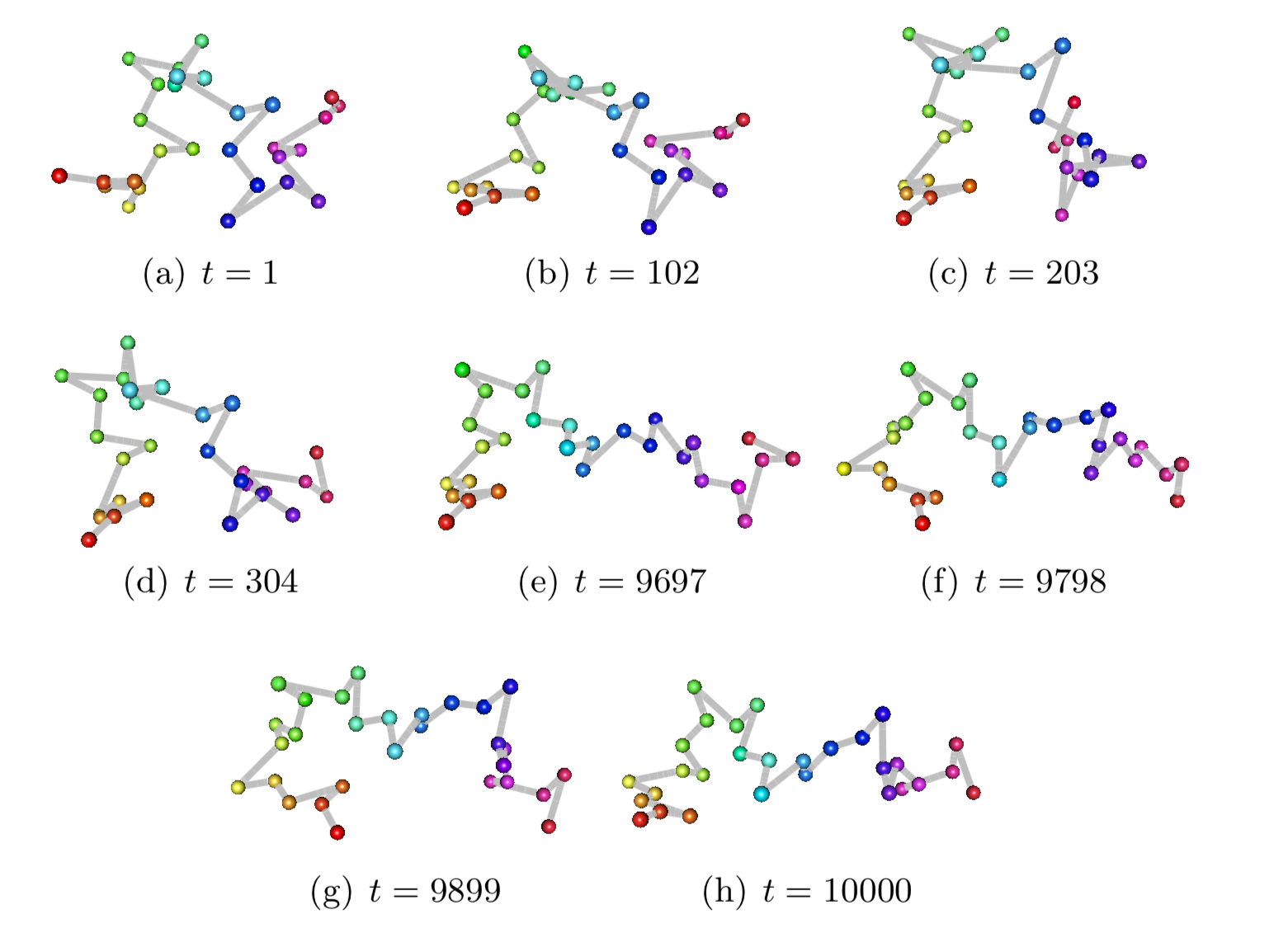}
 \caption{Temporal sequences of thinned run 1.} 
 \label{fig:moving}
  \end{figure}

The main motivation for this work comes from biomedical sciences, where 
it is of great interest to study the changes in shape of a protein, as its shape 
is an important component of a protein's function. Studies of this type are often  
approached through molecular dynamics simulations, which are large deterministic simulations using 
Newtonian mechanics to model the movement of a protein in a box of water molecules \citep[e.g.][]{AMBER13}. We consider a dataset of 100 independent simulation runs
in the study of the small alanine pentapeptide (Ala$_5$) which consists of $k = 29$ atoms in $\R^3$ at 
$10,000$ equal picosecond ($10^{-12}s$) time intervals. Further details on this particular peptide are given by \cite{Margulisetal02}. 
The $100$ runs all start with fairly similar (but different) configurations and  
the subsequent simulated configurations vary considerably over the $10000$ time points (10 nanoseconds). Figure \ref{fig:moving} shows example configurations at times near the beginning and end of the sequence for run 1, 
after removing the effect of rotation, translation and scaling using generalized 
Procrustes analysis \citep{Gower75,Goodall91}. 
As seen in the figures, the shape changes quite a lot with bending and straightening 
over the course of time.

Shape analysis can be considered an example
of Object Oriented Data Analysis \citep{Wangmarr07,Marralon14}, 
and the key initial question to ask is: ``What are the data 
objects?" We have several choices available in our application, 
for example the data objects could be (i) the shapes 
of each individual molecule of 29 atoms in 3 dimensions or (ii) individual runs of 
29 atoms in 3 dimensions observed at all 10,000 time points. The two respective choices of object space are (i) 
$\Sigma^{29}_3$ and (ii) $(\Sigma^{29}_3)^{10000}$, where $\Sigma^k_m$ is the shape space of 
$k$ points in $m$ dimensions \citep{Kendall84}. For the application in this paper we will consider (i) for our data objects, and so the full sample size is $n = 10,000 \times 100 = 1,000,000$ in our case, whereas in (ii) 
the sample size would be $n =100$.  


It is of interest to describe the shape variability using a low dimensional representation and 
to examine if there are preferred states, i.e. clusters of shapes which are more commonly formed by the 
dynamic peptide. Also, the patterns of 
temporal transitions between states are of interest. 

\section{Shape PCA, principal nested spheres and principal nested shape spaces}
\subsection{Shape PCA}
Consider $k \times m$ configuration matrices $\tilde{X}_i, i = 1, \ldots, n$, where $n$ is the number of configurations.
Shape PCA is similar to classical PCA \citep{Jolliffe02}, except the shapes must first be projected into the Procrustes 
tangent space centred at the overall estimated mean shape of the data. Technical details of the Procrustes tangent space are given in 
Appendix 1. The observations 
are optimally aligned (by translating, scaling and rotating) using generalized Procrustes analysis \citep{Gower75,Goodall91} 
and the full Procrustes sample mean is obtained as an estimate of the overall population mean shape 
\citep[Equation (6.11)]{Drydmard16}. The procedure is 
described in detail in Chapter 7 of \citet{Drydmard16} and is implemented in the {\tt shapes} package in R \citep{Dryden-shapes}. 
Let  $T_i, i = 1, \ldots, n,$ 
be the Procrustes tangent projection of centred, scaled and rotated 
configurations of $\tilde{X}_i$ at their Procrustes mean \citep[Equation (4.33)]{Drydmard16}, which 
are obtained using the command {\tt procGPA} with option {\tt tangentcoords="partial"} from the 
R package {\tt shapes}.  The principal components are the eigenvectors of the sample 
covariance matrix of $T_i, i = 1, \ldots, n$, and these eigenvectors and the PC scores are given in the output of {\tt procGPA}. 
The method is equivalent to carrying out PCA using the extrinsic Frobenius distance between matrices in the pre-shape space after removing rotations using Procrustes registration. 

\subsection{Principal nested spheres}\label{PNS}
The analysis of the principal nested spheres for a given data set in a unit sphere $\mathcal{S}^d$ is introduced in \cite{JDM12}. The main idea is that 
a high dimensional unit sphere is decomposed into successively lower dimensional subspheres using backwards
fitting, and at each level the Euclidean PNS scores are obtained from the residuals.  
The method can capture non-geodesic variation, so if the data lie on a curved submanifold of the sphere,
then the resulting variation after fitting the PNS can be linearly represented \citep{JDM12}.
PNS is an iterative method of decomposing $\mathcal{S}^d$ into a sequence $\{\mathcal{U}_{d-k}\mid k=1,\ldots,d-1\}$ of nested (sub-) spheres, where each $\mathcal{U}_i$ is a sphere of dimension $i$ and $\mathcal{U}_i\subset\mathcal{U}_{i+1}\subset\mathcal{S}^d$. At each step $k$, the method captures the variation of the data set in a lower dimensional subsphere and provides the best $(d-k)$-dimensional approximation $\mathcal{U}_{d-k}$ to the data. Since $\mathcal{U}_i$ is a great subsphere of $\mathcal{S}^d$ if and only if $\mathcal{U}_i$ is a unit sphere, the method extends the existing methods of finding principal geodesics for the data. The main ingredients of the method can be summarised as follows.

\vskip 6pt
Any subsphere of $\mathcal{S}^i$ of dimension $i-1$ can be characterised by $v\in\mathcal{S}^i$ and $r\in(0,\pi/2]$ as
\begin{eqnarray}
A_{i-1}(v,r)=\{x\in\mathcal{S}^i\mid \rho(v,x)=r\},
\label{eqn0}
\end{eqnarray}
where $\rho(x,y)$ is the spherical distance between $x,y\in\mathcal{S}^i$ given by $\rho(x,y)=\cos^{-1}\left(\langle x, y\rangle_{\mathbb{R}^{i+1}}\right)$. 

For a data set $\{x_1,\ldots,x_n\}\subset\mathcal{S}^d$, the best fitting $(d-1)$-dimensional subsphere is defined to be
\[\hat{\mathcal{U}}_{d-1}=A_{d-1}(\hat v_1,\hat r_1),\] 
where
\begin{eqnarray}
(\hat v_1,\hat r_1)=\argmin_{v_1\in\mathcal{S}^d,r_1\in(0,\pi/2]}\sum_{i = 1}^n\epsilon_{i,d-1}(v_1, r_1)^2
\label{eqn1}
\end{eqnarray}
and $\epsilon_{i,d-1}(v_1,r_1)=\rho(x_i,v_1)-r_1$. Then 
\[E(d-1)=(\hat\epsilon_{1,d-1},\ldots,\hat\epsilon_{n,d-1})^\top\]
are the signed residuals on the sphere, where $\hat\epsilon_{i,d-1}=\epsilon_{i,d-1}(\hat v_1,\hat r_1)$. Since $\hat{\mathcal{U}}_{d-1}$ has radius $\sin(\hat r_1)$, denote by $x^p_i$ the projection of $x_i$ onto $\hat{\mathcal{U}}_{d-1}$ divided by $\sin(\hat r_1)$, so that $\{x^p_1,\ldots,x^p_n\}\subset\frac{1}{\sin(\hat r_1)}\hat{\mathcal{U}}_{d-1}$. Since $\frac{1}{\sin(\hat r_1)}\hat{\mathcal{U}}_{d-1}$ is isometric with the standard unit sphere $\mathcal{S}^{d-1}$, applying the above procedure to $\{x^p_1,\ldots,x^p_n\}$ we get the best fitting $(d-2)$-dimensional (sub-)sphere to the data to be
\[\hat{\mathcal{U}}_{d-2}=\sin(\hat r_1)\,A_{d-2}(\hat v_2,\hat r_2)=\{\sin(\hat r_1)\,x\mid x\in A_{d-2}(\hat v_2,\hat r_2)\}\subset\hat{\mathcal{U}}_{d-1},\]
where $(\hat v_2,\hat r_2)$ is determined by \eqref{eqn1} using $\{x^p_1,\ldots,x^p_n\}$ with $d$ replaced by $d-1$ and $\mathcal{S}^d$ replaced by the $(d-1)$-dimensional unit sphere $\frac{1}{\sin(\hat r)}\,\hat{\mathcal{U}}_{d-1}$, and where $A_{d-1}(\hat v_2,\hat r_2)$ is defined on $\frac{1}{\sin(\hat r)}\,\hat{\mathcal{U}}_{d-1}$ in a similar fashion to that in \eqref{eqn0}. Moreover, the corresponding residuals are
\[E(d-2)=\sin(\hat r_1)\,(\hat\epsilon_{1,d-2},\ldots,\hat\epsilon_{n,d-2})^\top.\] 
The resulting fitting sequence $\{\hat{\mathcal{U}}_{d-k}\mid k=1,\ldots,d-1\}$ obtained by repeating this procedure is the so called principal nested spheres for the given data set $\{x_1,\ldots,x_n\}$. Writing $E(0)$ for the spherical coordinates of the final projection points in $\hat{\mathcal{U}}_1$ of the data, with respect to their Fr\'echet mean in that sphere (assuming it is unique), the resulting combined residuals
\[\begin{bmatrix}E(0)&E(1)&\cdots&E(d - 1)\end{bmatrix}^\top\]
give a representation of the data, where the $i$th column comprises the coordinates of $x_i$ in terms of the principal nested spheres. This representation is called the PNS coordinates of the data and can be used to interpret the structure of the data.

\subsection{Principal nested shape-spaces} 
Principal nested spheres analysis for a data set on a sphere uses the particular structure of the sphere. In principle, this can be generalised to a sequence of fitting nested sub-manifolds for a data set contained in a manifold. However, it is generally not straightforward. For the analysis of shape variation of a given set of configurations, we propose a method of applying the technique of PNS to obtain principal nested sub-shape spaces (PNSS) for shapes of $m$-dimensional configurations. 

We wish to describe the shape variability of the peptide data, and so this requires removing information 
about translation, scale and rotation.
For a given configuration $\tilde X$ in $\mathbb{R}^m$ with $k > m$ labelled landmarks that are not all identical, the pre-shape $X$ of the configuration is obtained from $\tilde X$ by removing the effects of translation and scaling. The pre-shape $X$ can be represented by a $(k-1)\times m$ matrix of unit norm \citep{Kendall84} and it is known that the pre-shape sphere $\mathcal{S}_m^k$ consisting of all pre-shapes of such configurations is the entire sphere $S^{m(k-1)-1}$. Then, the shape $[X]$ of $\tilde X$ is the equivalence class of $X$ under rotation, i.e. $[X]=\{XR\mid R\in SO(m)\}$.

In order to remove rotations we consider the quotient space of the 
pre-shape sphere $\mathcal{S}_m^k$ with respect to rotations $SO(m)$, and this shape space 
is complicated for $m \ge 3$ being non-homogenous and containing 
singularities \citep{Lekend93}. However, practical statistical analysis can be carried out 
by identifying the tangent space ${\mathcal T}_{X_0}(S^k_m)$ to the pre-shape sphere 
at a point $X_0$ as comprising 
two orthogonal real sub-spaces: the vertical and horizontal tangent spaces. 
The vertical tangent space ${\mathcal V}_{X_0}$ contains the rotation information and the horizontal
tangent space ${\mathcal H}_{X_0}$ contains the shape information, and the latter is often called the
Procrustes tangent space \citep{Kentmard01}.  Working in this horizontal tangent space forms the heart of
most practical statistical analyses of landmark shapes.  

To develop the method of PNSS we need to make some further identifications. We consider 
the subset of the pre-shape sphere $S^k_m$ which is orthogonal to ${\mathcal V}_{X_0}$, 
and denote it by $S_{X_0}$ which a great sphere of dimension $(k-1)m-m(m-1)/2-1$. 
In fact $S_{X_0}$ is the image of ${\mathcal H}_{X_0}$ using the exponential map onto the sphere. 
If we now have a datapoint $X \ne X_0$ on the pre-shape sphere then we can obtain the 
Procrustes fit as the solution to minimizing the great circle distance $\rho(XR,X_0)$ over rotations $R$. The Procrustes 
fit is denoted by $S_X = XR_X$ with the fitted rotation matrix $R_X$, and for all $X \in \mathcal{S}_{X_0}$ the 
Procrustes fits $S_X$ lie within 
a half-sphere of $\mathcal S_{X_0}$, because the distance $\rho(S_X,X_0) \le \pi/2$. The Procrustes fit will be 
unique if the rank of $X$ is at least $m-1$ and $[X]$ is outside the cut-locus of $[X_0]$. In the case of uniqueness,  
which will often be the case for practical data, we can identify the subset of the sphere
\[\mathcal{B}_{X_0}=\{S_X\in\mathcal{S}_{X_0}\mid X\in \mathcal{S}_m^k\hbox{ and }S_X\hbox{ is unique}\}\] 
with a bijective map with the subset of shape space
\[\mathcal{B}_{[X_0]}=\{[X]\in\Sigma_m^k\mid[X]\hbox{ is non-singular and not in the cut locus of }[X_0]\}.\] 
The subsets $\mathcal{B}_{X_0}$ and $\mathcal{B}_{[X_0]}$ are diffeomorphic to each other. The practical implication of this identification is that any sub-manifolds in $\mathcal{B}_{X_0}$ are mapped to sub-manifolds in $\mathcal{B}_{[X_0]}$.\\

{\bf Definition:} 
{\it Given a sequence of principal nested spheres on ${\mathcal S}_{X_0}$ (as defined in Section \ref{PNS}), 
the intersection of the sequence of principal nested spheres with $\mathcal{B}_{X_0}$  maps to a sequence of 
sub-manifolds in $\mathcal{B}_{[X_0]}$, which we define as a sequence of {\bf Principal Nested Shape Spaces}.  The final point in the sequence, with dimension 0, corresponds to the {\bf PNSS mean shape}. }\\

We now consider the practical implementation of principal nested shape spaces for the peptide shape data in $m=3$ dimensions. 
First of all an overall Procrustes mean $\bar X$ is obtained using 
generalized Procrustes analysis\nocite{Gower75,Drydmard16}. We use the function {\tt procGPA} in the {\tt shapes} package
in R \citep{Dryden-shapes}. For our dataset each Procrustes fit is unique, as will generally be the case for 
most practical datasets. 
We then apply PNS to the Procrustes fitted data on $S_{\bar X}$, and the resulting PNS subspaces intersections
with $\mathcal{B}_{\bar X}$ will be have 
a diffeomorphic mapping to the PNSS subspaces in $\mathcal{B}_{[\bar X]}$, 
and the PNS scores are equal to the PNSS scores.    

PNSS uses the relationship between the pre-shape sphere 
$\mathcal{S}_m^k$ and the shape space $\Sigma_m^k$ to construct a nested sequence of sub-shape-spaces in a given shape space from a sequence of nested subspheres, as constructed in the previous section, in the corresponding pre-shape sphere. Despite not being strictly analogous to the concept of principal nested spheres, this approach does offer an extrinsic method to apply that concept. The $m=2$ dimensional case was discussed by \cite{JDM12}, where 
complex arithmetic leads to a simple adaptation of PNS to planar shapes, where the shape space 
is the complex projective space ${\mathbb C}P^{k-2}$, which is a homogeneous space \citep{Kendall84}. 

A full technical construction of PNSS is given in Appendix 1.  

\subsection{Approximate PNSS using PC scores}
When a dataset is large, the numerical computation required for the principal nested spheres analysis, and so for the principal nested sub-shape-spaces analysis, usually tends to be slow. In fact the time required to compute PNSS on large datasets can be extensive. Much of the time is spent fitting the higher dimensional spheres, which are taking account of small amounts of noise. An alternative approximate approach is to first carry out PCA and retain the first $p$ PC
scores, where these scores capture a high percentage of the variability. 
If the Procrustes fits of the original data configurations to their mean shape $\bar X$ are close to a great subsphere of a lower dimension than the dimension of $\mathcal{B}_{\bar X}$, the speed of computation can be much improved by first using the technique of principal component analysis to determine this great subsphere and then using the projections to this subsphere as the input for the principal nested sub-shape-spaces analysis.  

The technical details of the method of approximate PNSS using PC scores are given in Appendix 2.

A practical implementation of PNSS using PC scores is given in the {\tt shapes} package in 
R using the command {\tt pns4pc} \citep{Dryden-shapes}, and this code is used in our peptide application.  
Output from the R command {\tt pns4pc} includes the PNSS scores, where the first PNSS score is a circular variable and the remaining 
PNSS scores are Euclidean. In our application we shall see that the clusters of peptide states are much more apparent in the PNSS scores
compared to the PC scores.  The percentage of shape variability explained by the PNSS scores is helpful to explore the 
efficiency and effectiveness of the method, and we will see that the first few PNSS scores explain much more 
shape variability than the first few PC scores. 

\section{Peptide shape analysis}
\subsection{Molecular dynamics data}
The data set consists of a total of $100$ runs, with each run consisting of 
$k = 29$ landmarks in $m=3$ dimensional space from a small peptide (Ala$_5$)
evaluated  at consecutive $n=10000$ times at picosecond intervals. 
As the time interval between consecutive configurations is very small, we first thin the 
data in time to give $100$ configurations at equally spaced times from each run. 
The thinned times are $t_1 = 1, t_2 =102, t_3 = 203,\ldots,t_{99}=9899, t_{100}=10000$.
We consider temporal sequences of each run and register using generalized 
Procrustes analysis \citep{Gower75,Goodall91}.

\begin{figure}[htbp]
    \centering \subfigure[Histograms]
    {\includegraphics[scale = .40]{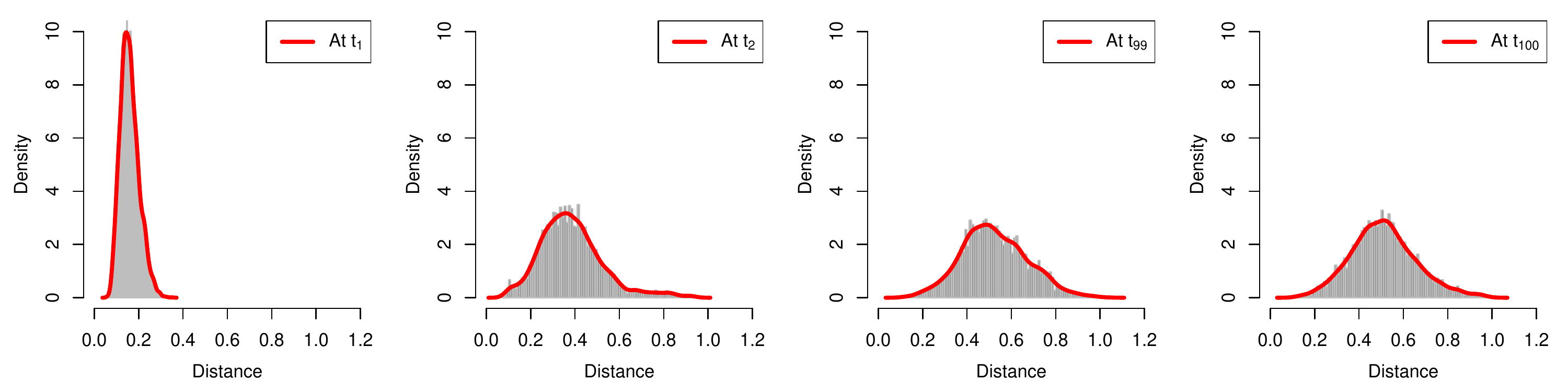}}
    \centering \subfigure[Boxplots]
    {\includegraphics[scale = 0.38]{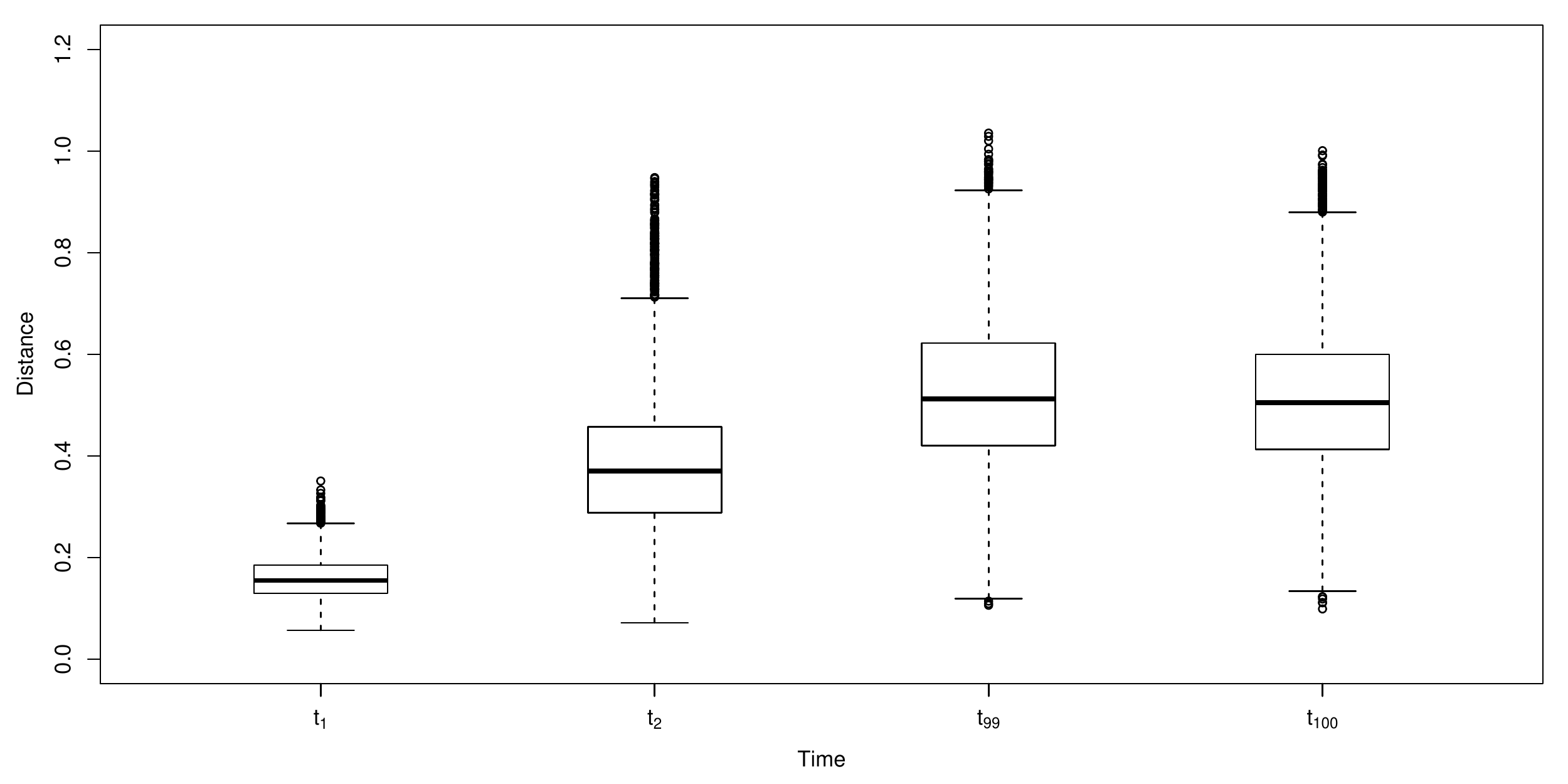}}
    \caption{Histograms (a) and boxplots (b) of pairwise Riemannian shape distances at first two starting and last two finish times.
    For $i = 1, 2, 99, 100$, $\rho \big( \big[ \tilde X^{(r_1)}_{t_i} \big], \big[ \tilde X^{(r_2)}_{t_i} \big] \big), r_1, r_2 = 1, 2, \ldots, 100$.}
    \label{fig:RDist.histbox}
\end{figure}

The starting configuration for each peptide is quite similar 
at the start of each run with the 
Riemannian shape distance $\rho$ between pairs of runs primarily less than $0.3$ radians, where $0 \le \rho \le \frac{\pi}{2}$. The variability in shape between runs is much 
larger at the end (Riemannian distances up to about $1.1$), as see in Figure  \ref{fig:RDist.histbox}. 

\subsection{Shape PCA and PNSS}
\begin{figure}[htbp]
    \centering
    {\includegraphics[scale = 0.27]{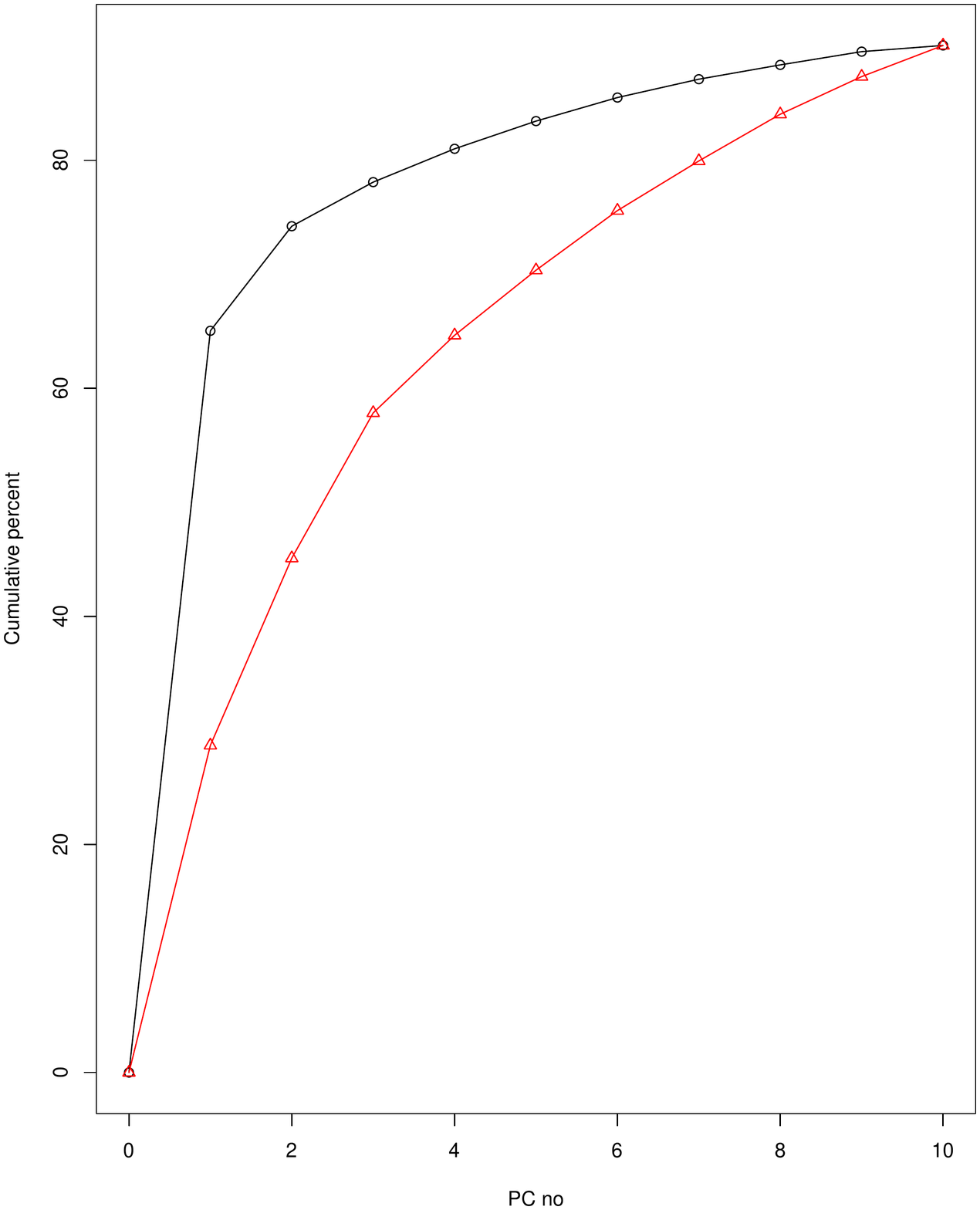}}
   {\includegraphics[scale = 0.27]{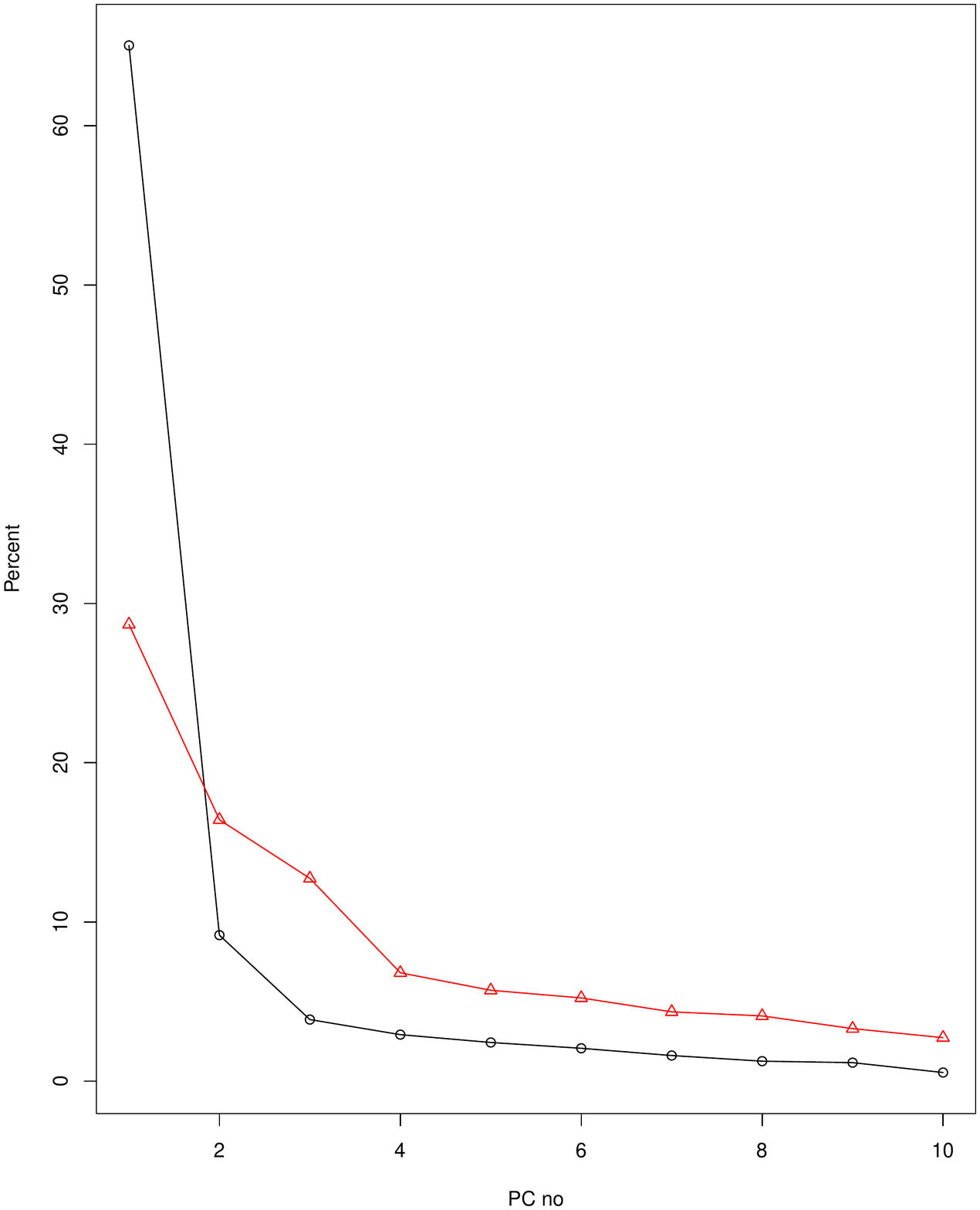}}
    \caption{(Left) Cumulative percent explained by principal components for PNSS (black) and shape PCA (red). 
 (Right) The percentages of shape variability captured by the individual PC scores  for PNSS (black) and shape PCA (red).}
    \label{fig:eig}
\end{figure}

We first run shape PCA on the thinned 100 configurations from each of 100 runs (10000 configurations in total),
Much of the data variation can be much explained by first several principal components as shown 
in Figure \ref{fig:eig}.
We choose the first ten shape principal components for the computation of PNSS using PC scores 
since those ten explain 90.1\% of the shape variation. The percentages of shape variability captured by the first three individual PC scores are 
28.7\%, 16.4\%, 12.7\%  and by the PNSS scores are 65\%, 9.2\%, 3.9\%. 

We carry out the PNSS analysis to sequentially reduce the dimension of the 
projected data from $S^{10}$ to $S^1$ to $S^0$. The percentage variability 
explained by the first two PNSS scores ($74.2\%$) is much higher than the first two PCA scores ($45.1\%$). 
The PNSS score 1 is a circular variable, where the most negative score and most positive PNSS1 score are identified with each other 
(at the cut point of the circle). The higher PNSS scores do not have the extreme points identified with each other.



\begin{figure}[htbp]
   \centering \subfigure[PCA plot.]
   {\includegraphics[scale = 0.45]{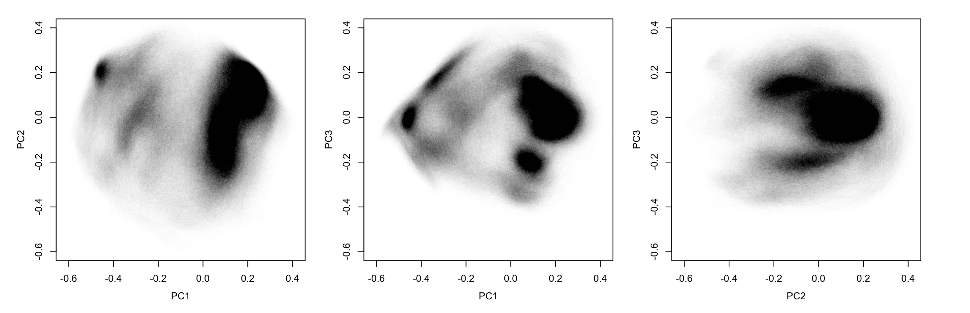}}
   \centering \subfigure[PNSS plot using 10 shape PC scores.]
   {\includegraphics[scale = 0.45]{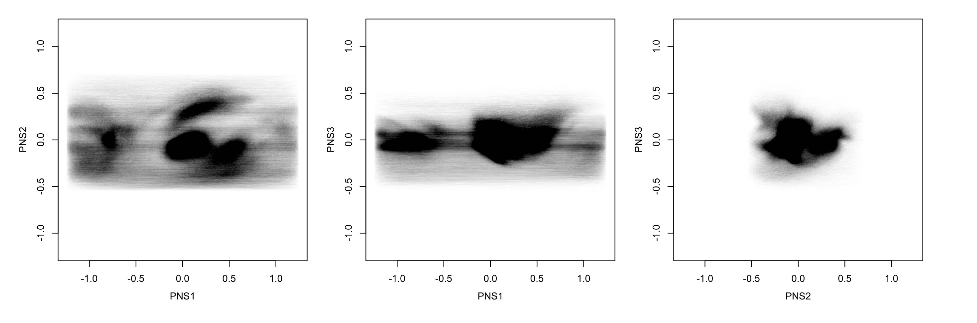}}
   \caption{Pairwise components plot of (a) shape PCA and (b) PNSS using first 10 shape PC scores.}
   \label{fig:pc.pnspc}
\end{figure}

We calculate PNSS1, PNSS2 and PNSS3 coordinates of all the 1,000,000 configurations
through the model constructed using the thinned 10000 configurations.
The pairwise 2D plots of first three PNSS and PC scores are in Figure \ref{fig:pc.pnspc}.
A vertical long cluster forms all over PC2 between small range of PC1.
On the other hand, in PNSS1-PNSS2 plot
three salient clusters are close to each other and one cluster is positioned on the left side.
For this particular peptide there are believed to be four preferred shape states, and these can be clearly 
seen in the first PNSS plot, whereas it is not evident with PCA.  Hence, there is a clear advantage in 
using PNSS compared to PCA in this dataset.

The relative variation of PNSS1 to PNSS2 is much higher than that of PC1 to PC2.
Since first two PNSS components account for high proportion (74.2\%) of the shape variability,
the subsequent components from PNSS3 are far less important.
Both left and right tails in PNSS1 should be considered connected, given this is a circular variable.


\subsection{Cluster analysis}
\begin{figure}[htbp]
   \centering \subfigure[Clustering on (PC1, PC2, PC3).]
   {\includegraphics[scale = 0.4]{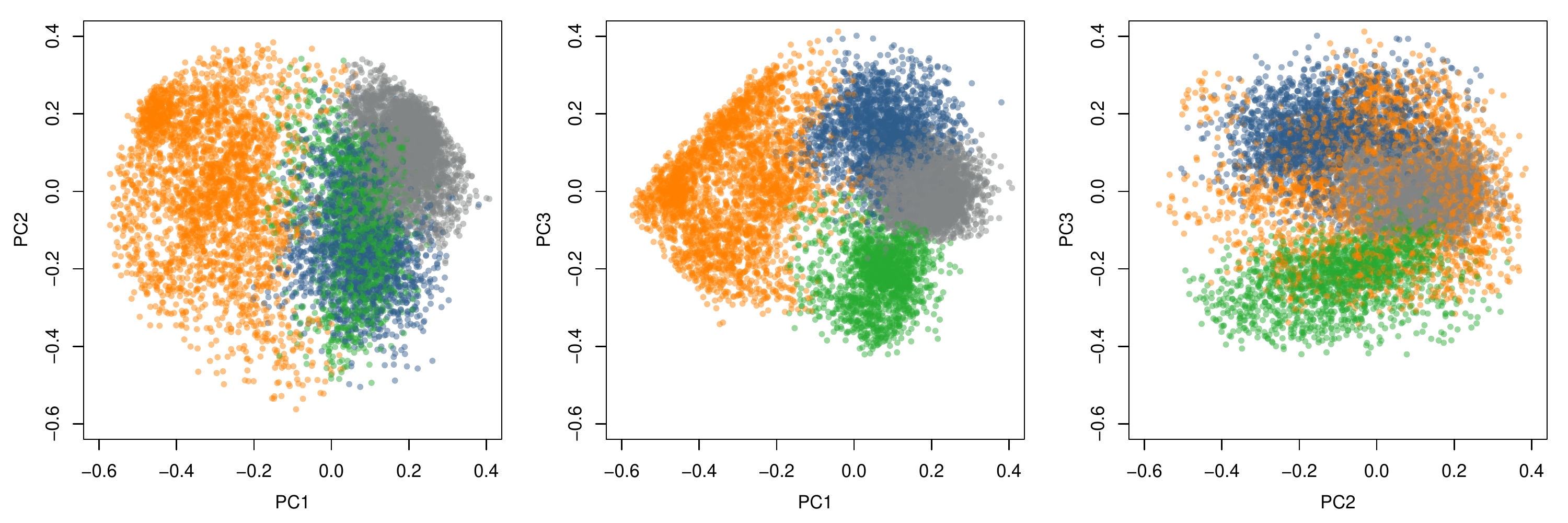}}
   \centering \subfigure[Clustering on $S^{10}$ data and representation in terms of (PNSS1, PNSS2). PNSS1 is a circular variable, with $-1.246$ and $1.246$ being equivalent, as the cut-point of the circle. ]
   {\includegraphics[scale = 0.32]{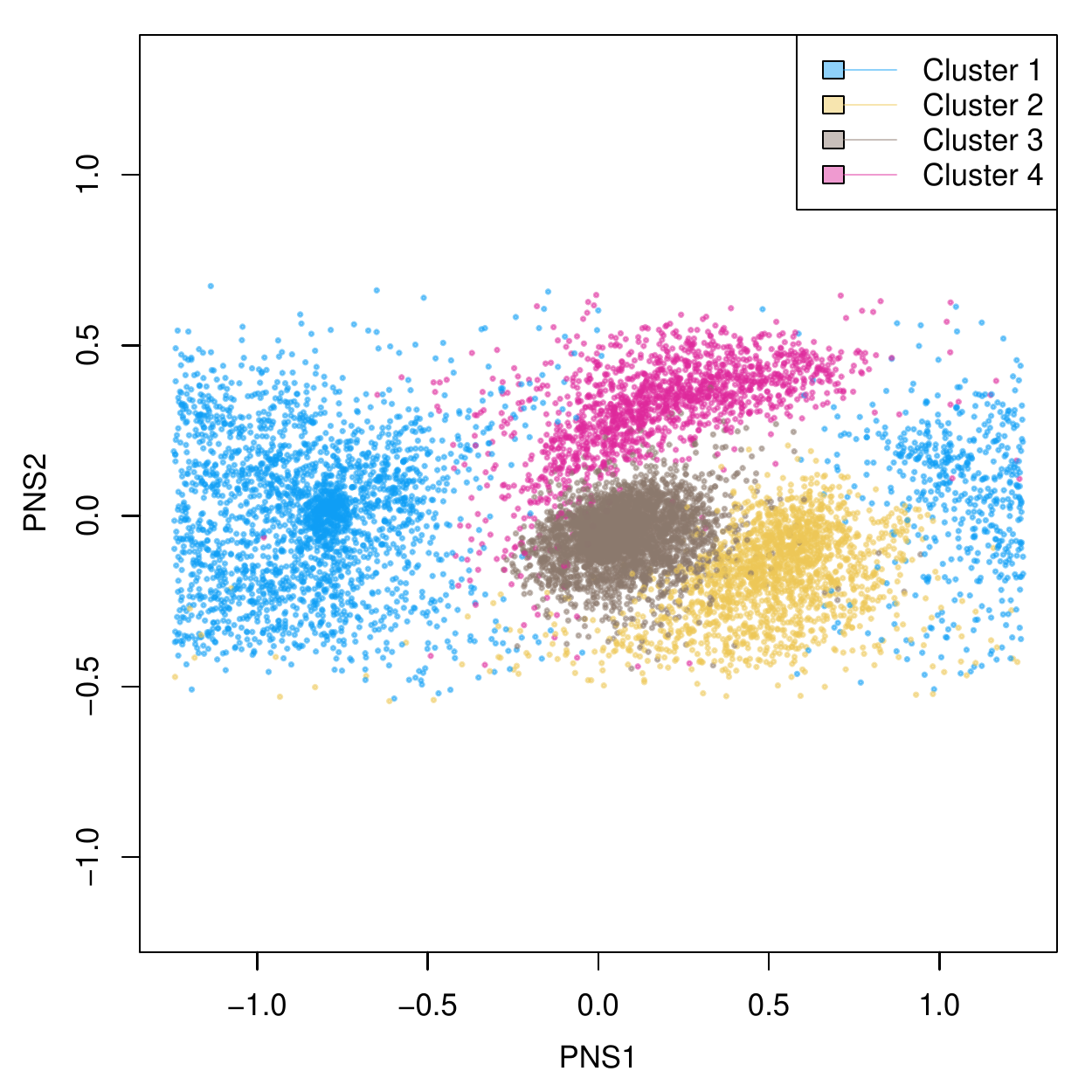}} $\qquad $
   \centering \subfigure[Clustering on (PC1, PC2, PC3) and representation in terms of (PNSS1, PNSS2). PNSS1 is a circular variable, with $-1.246$ and $1.246$ being equivalent, as the cut-point of the circle. ]
   {\includegraphics[scale = 0.32]{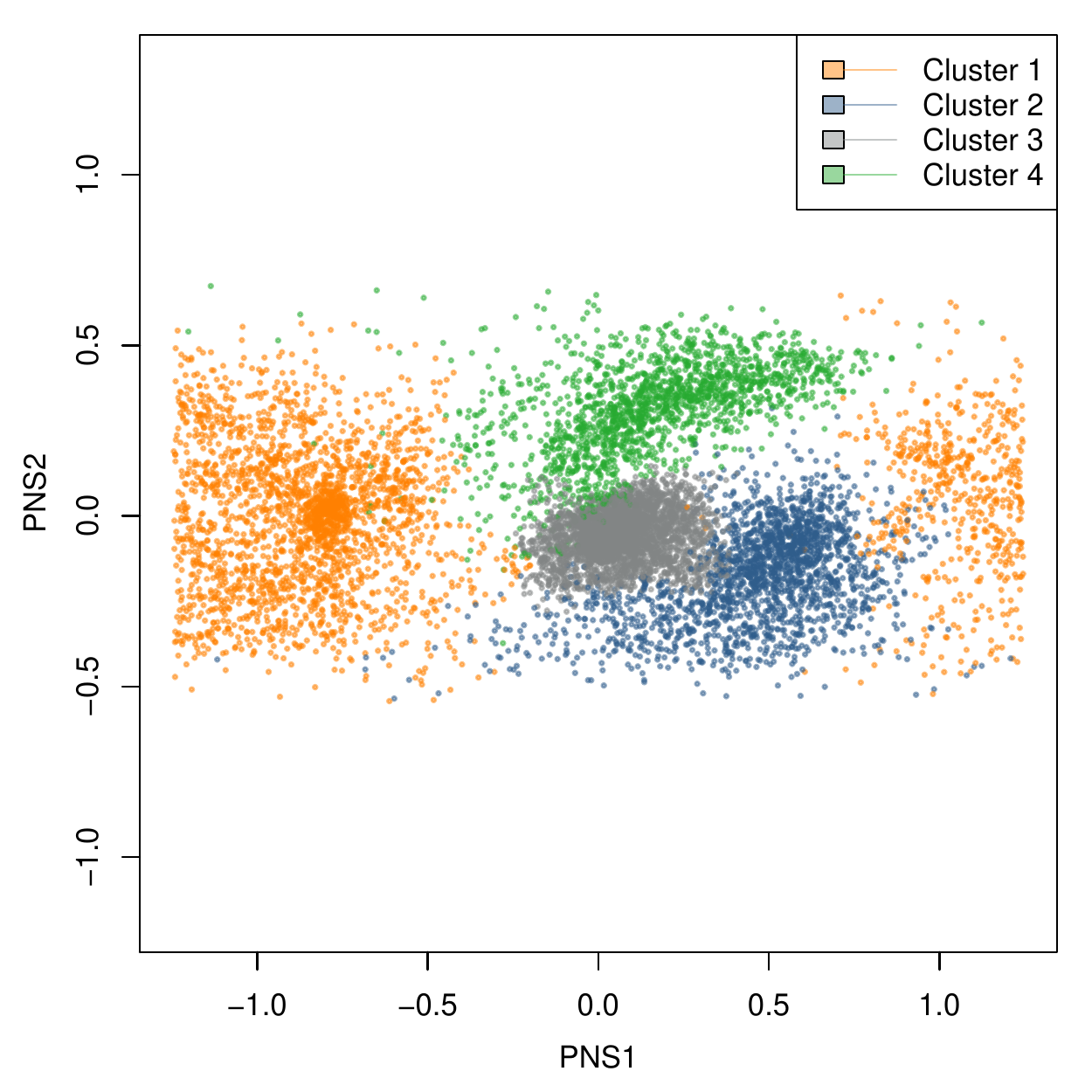}} \\
    \centering \subfigure[$S^{2}$ data. PNSS clusters.]
{\includegraphics[scale = 0.225]{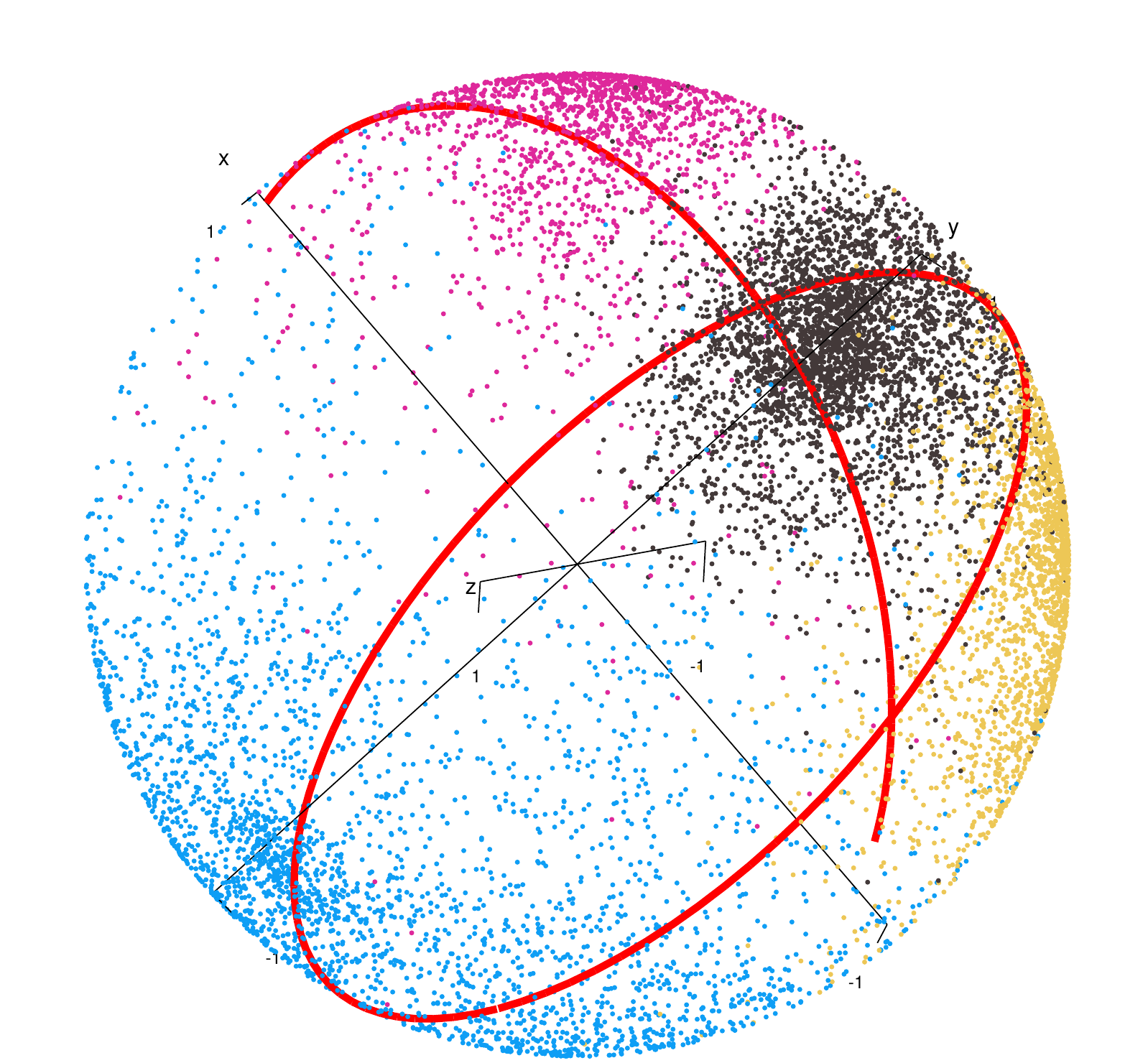}} $\qquad$
    \centering \subfigure[$S^{2}$ data. PCA clusters.]
   {\includegraphics[scale = 0.225]{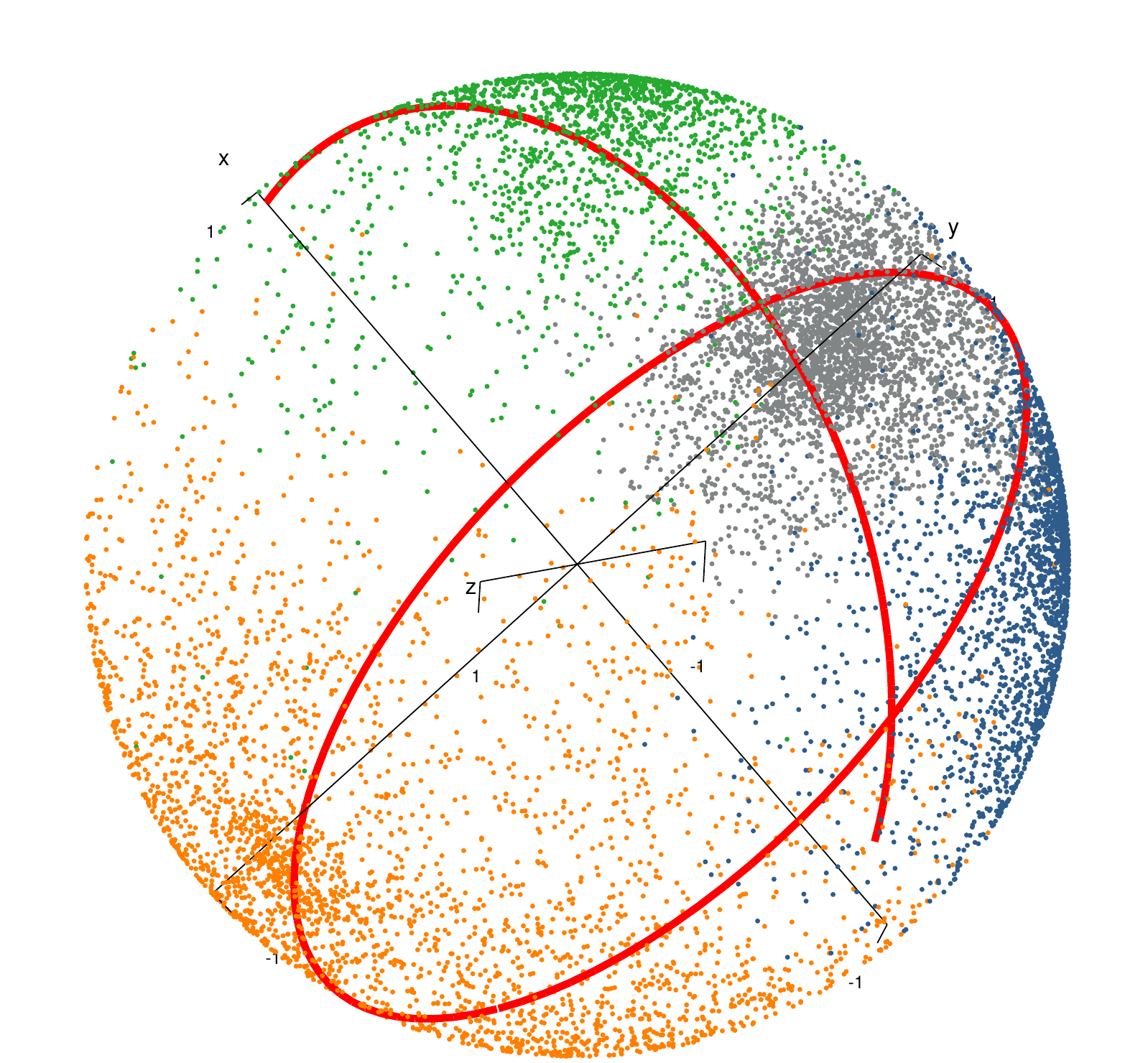}}
%
   \caption{Pairwise plots of PC1, PC2, PC3 scores  (a) and PNSS1, PNSS2 scores (b)-(e). The colour is using 
four-group hierarchical clustering on (PC1, PC2, PC3) in (a), (c) and (e) and on $S^{10}$ in (b) and (d), 
all using Ward's method with non-squared distances.
 Panel  (d), (e) shows the projected data points on $S^2$ with the estimated first two principal arcs. }
   \label{fig:clust}
\end{figure}

Given that there are believed to be four preferred states we now consider cluster analysis.
The result of four-group clustering is shown in Figure \ref{fig:clust},
where hierarchical clustering analysis using Ward's method is used \citep{Ward63} but with distances rather than squared 
distances ({\tt ward.D} in R).
In (a) the cluster analysis was performed on (PC1, PC2, PC3)
and in (b) it was performed using the great circle distance on $S^{10}$ then 
displayed in terms of (PNSS1, PNSS2) coordinates.
Three PCs in (a) are needed for separating them into four groups,
on the other hand two PNSSs in (b) are enough to split them clearly. However when looking carefully at 
Figure \ref{fig:pc.pnspc} the four clusters do not stand out in the PCA plot even with three PCs, but 
they are clearly distinct in the PNSS1-PNSS2 plot. 
Plotting PNSS1 and PNSS2 with the colour scheme from PCA clustering is given in (c), 
and we see that the result is similar. 
The most dense part of the blue cluster in (b) is where the starting configurations 
for each of the runs are located. Ward's method was our preferred clustering method, giving good
practical separation into four groups in our application. 

In (d) we plot the PNSS scores on $S^2$ using the PNSS clustering.  
The solid red line is the estimated first principal arc,
on the other hand the solid-dotted red line is the estimated second principal arc
where the solid line indicates the PNSS2 values. In (e) we plot the same data as in (d) except we
use the PCA cluster colours, and the clustering in the plots (d) and (e) are very similar. 
If we carry out the clustering on other sets of variables such as PNSS1, 2 then the broad pattern is 
similar, particularly near the main modes for the four clusters.

\subsection{Structure of PNSS variability}
For each PNSS we can obtain an interval on $S^{10}$ as
\begin{equation}\label{eq:interval}
\hat\mu \pm c s_j e_j,
\quad
j = 1, \ldots, 10,
\end{equation}
where $\hat\mu$ is the PNSS mean, $c$ is a constant, $s_j$ are standard deviation of $j$th PNSS scores
and $e_j$ are the direction of $j$th principal arc.
As seen in Figure \ref{fig:eig},
$s_1 = 0.5675, s_2 = 0.2131, s_3 = 0.1384$ and we used $c = 1$.

\begin{figure}[htbp]
   \centering 
      {\includegraphics[height=4.4cm]{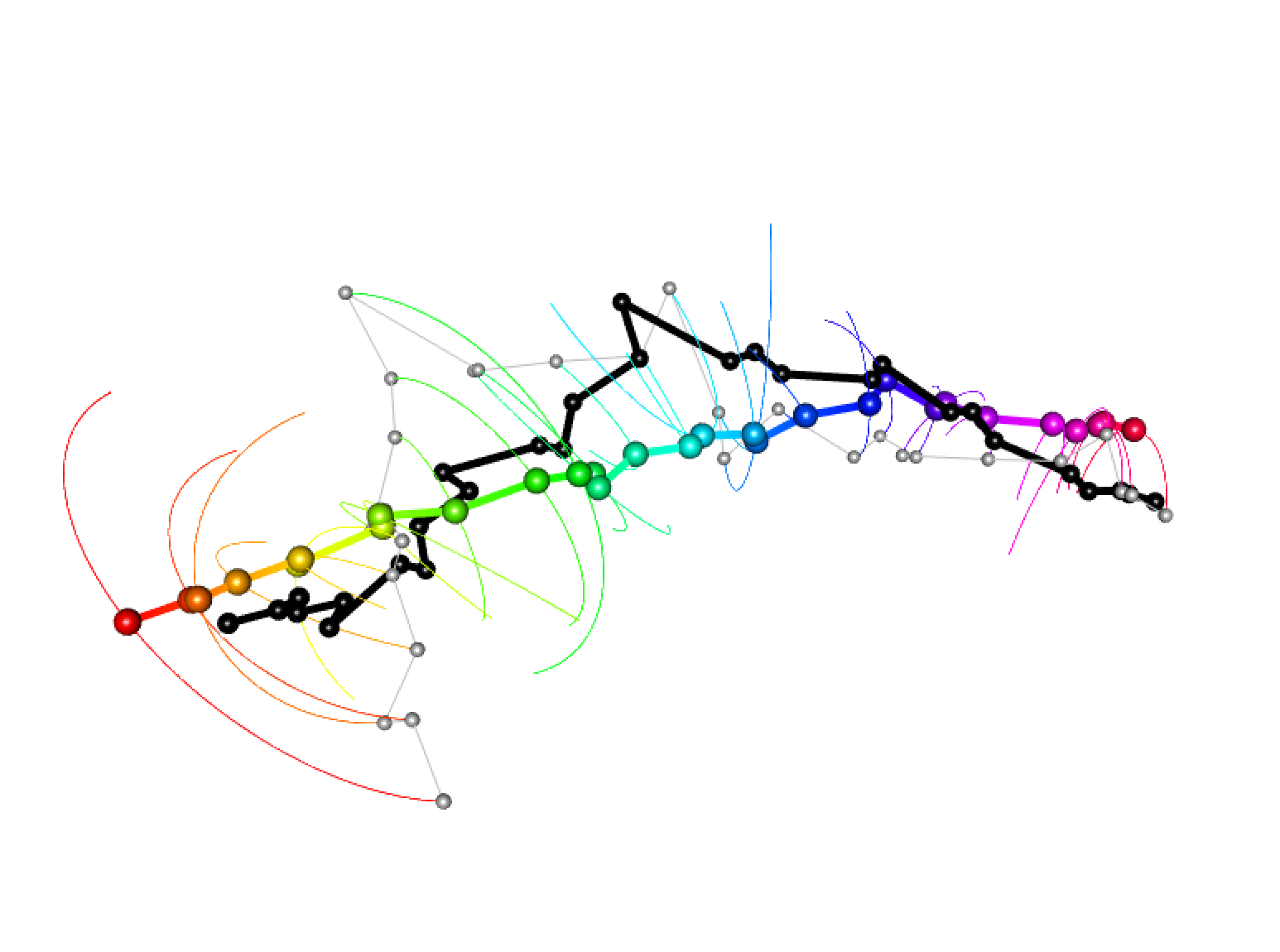}}
     {\includegraphics[height=3.3cm]{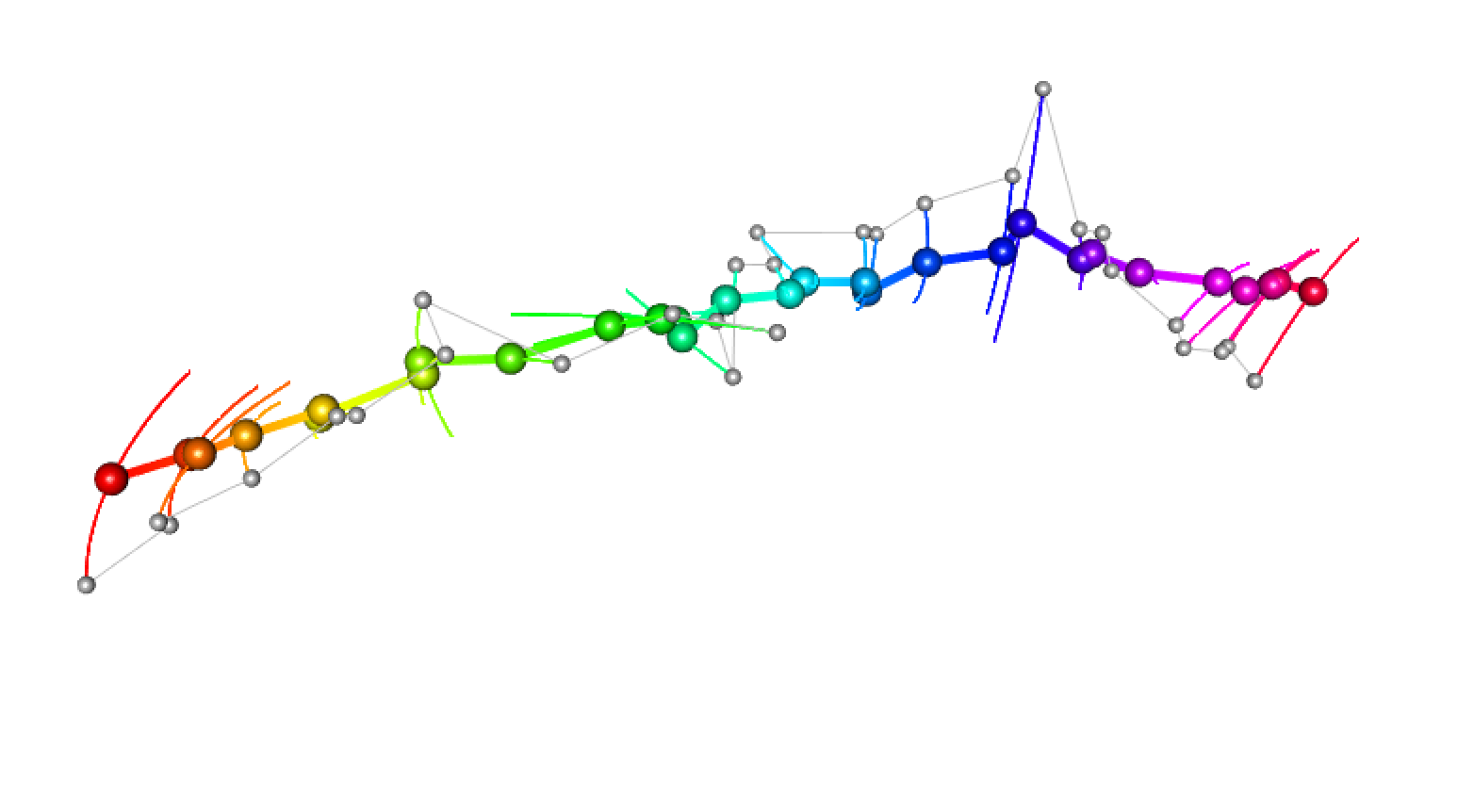}}
   {\includegraphics[height=2.75cm]{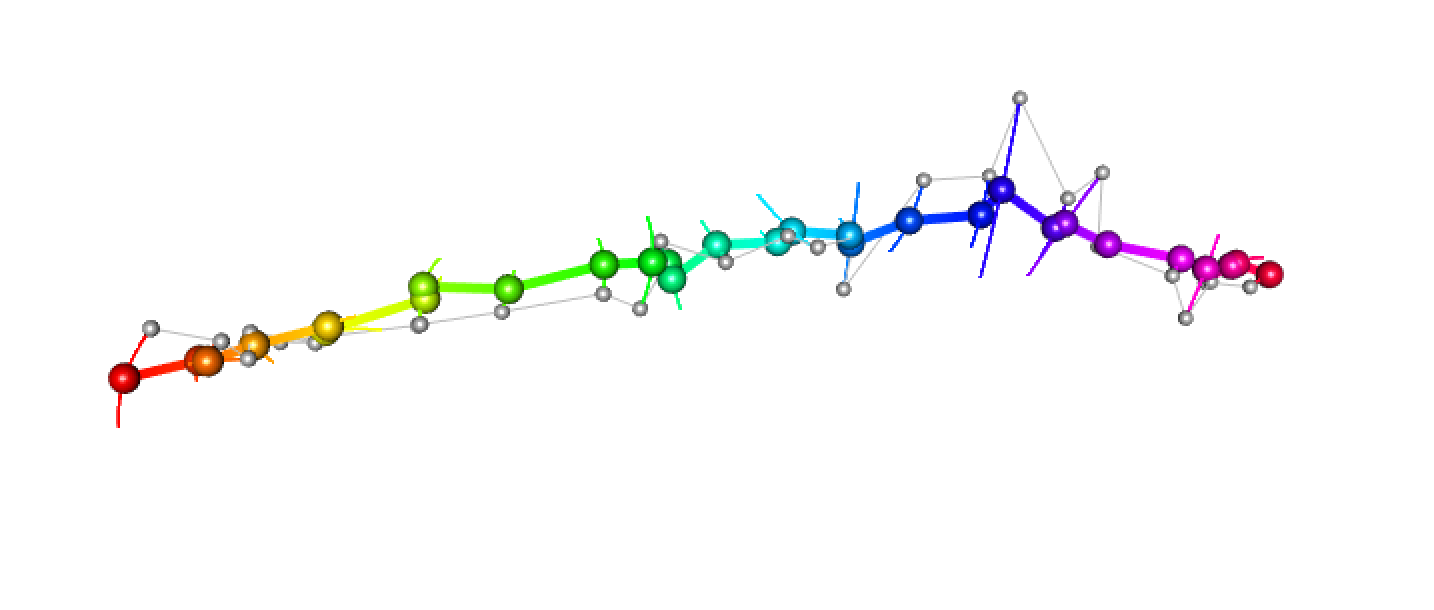}}
     \caption{PNSS 1 (top left), PNSS 2 (top right) and PNSS 3 (bottom) in landmark space, with PNSS mean (rainbow coloured 
points) on all plots, and the Procrustes mean (black) shown on the PNSS1 plot. Rainbow arcs have been drawn from the PNSS mean along each PNSS score direction to the ends of the interval (\ref{eq:interval}). 
The grey points and lines show the landmarks at $c=1$. }   
   \label{fig:arcs1}
 \end{figure}

We display the first three principal arcs in landmark space
as shown in Figure \ref{fig:arcs1} and we display both the PNSS mean (corresponding to PNSS component 0)  and the Procrustes mean for comparison. 
The black dots with connected black lines indicate the Procrustes mean. The PNSS mean 
is given by rainbow coloured points (with the rainbow colour indicating point order). We can see that the PNSS mean and Procrustes mean shapes are visibly different. 
In addition coloured rainbow arcs have been drawn from the PNSS mean along the PNSS score 1 direction to the ends of the interval (\ref{eq:interval}). 
The grey points indicate the location at $c=1$ along the arc, and these are joined by grey lines which give an idea of the structure of variability in the PNSS. 
This plot shows considerable amounts of twisting along the arcs in PNSS1, and a smaller amount of different bending in PNSS2 and PNSS3.

We see that the first PNSS score demonstrates an articulated type of movement, sweeping out 
arcs. The PNSS mean is clearly different from the Procrustes mean. The PNSS description of variation
is particularly appropriate here as bonds have fixed lengths, and local rotational movement is
much more practical than linear movement to and from the Procrustes mean, as in shape PCA.



\begin{figure}[htbp]
   \centering 
      {\includegraphics[height=3.6cm]{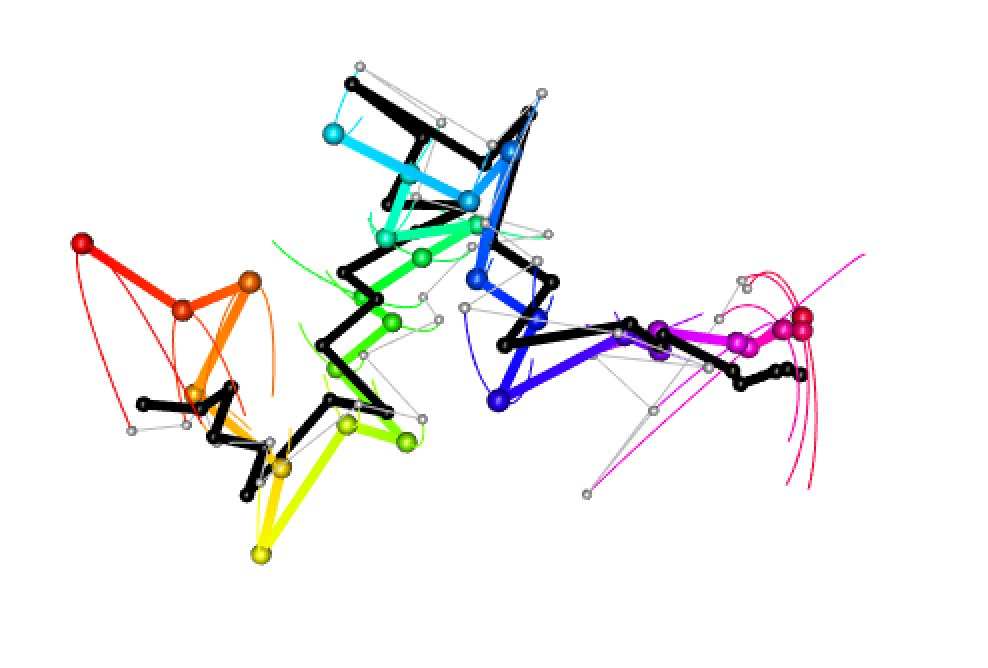}}
         {\includegraphics[height=3.3cm]{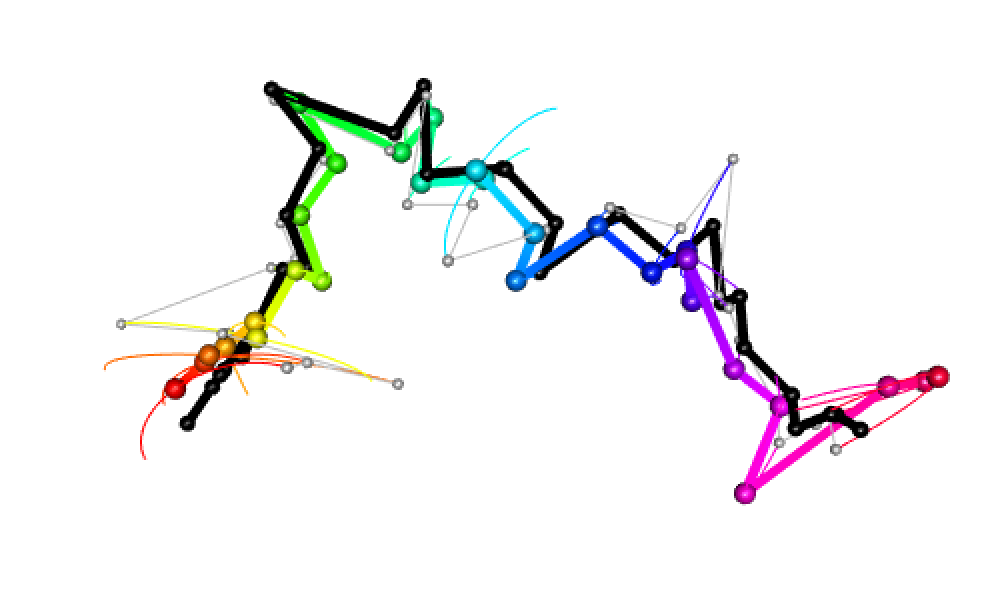}}
       {\includegraphics[height=3.6cm]{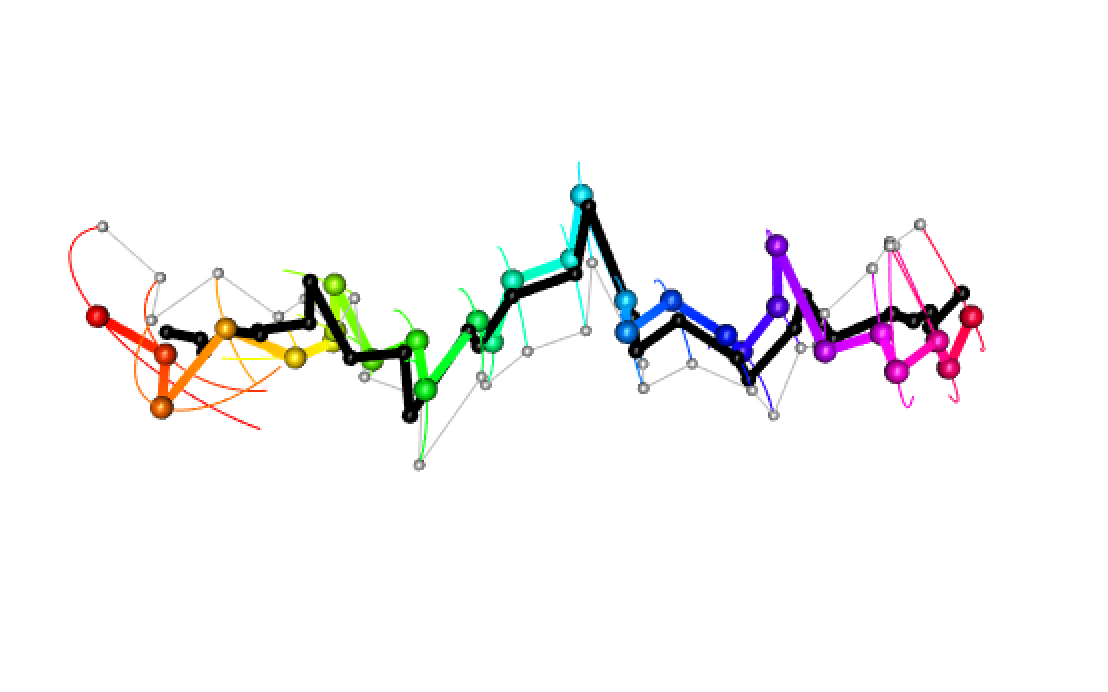}}
    {\includegraphics[height=3.3cm]{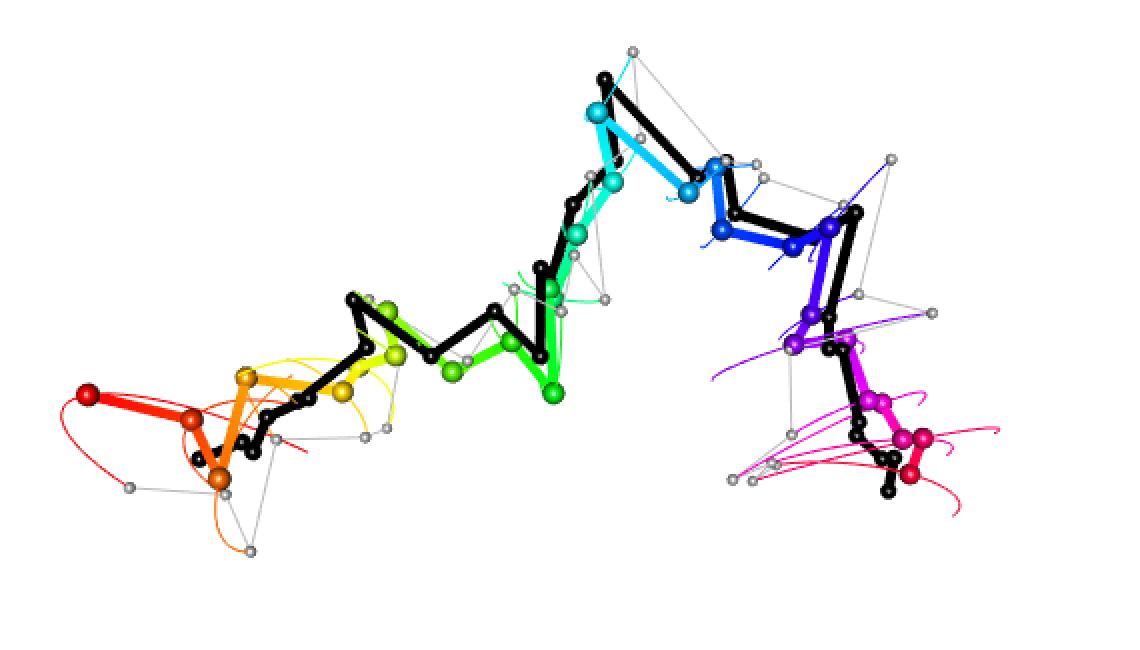}}
   \caption{PNSS 1 for each of the clusters 1 and 2 (top row, left, right) and clusters 3 and 4 (bottom row, left, right). The PNSS mean (rainbow coloured 
points) and Procrustes means (black) are displayed. Rainbow arcs have been drawn from the PNSS mean along the PNSS 1 score direction to the ends of the interval (\ref{eq:interval})}
   \label{fig:arcs2}
\end{figure}

In order to examine the shapes of the clusters, we plot the PNS mean and Procrustes shapes of each cluster in 
Figure \ref{fig:arcs2}, together with the PNSS 1 arcs for each cluster. The cluster 1 mean has two clear bends in the peptide, clusters 2 and 4 means 
have a single bend each in different positions, and cluster 3 mean is broadly straight with no bends. These four 
cluster means are quite different, 
and they illustrate the wide range of shape configurations that the peptides visit during their temporal sequences. Also, cluster 1 is clearly more variable, with larger 
PNSS 1 arcs than the other clusters. Also, the PNSS mean and Procrustes mean are more different in cluster 1. 


\subsection{Temporal clustering}
\begin{figure}[htbp]
   \centering \subfigure[Run 55]
   {\includegraphics[scale = 0.4]{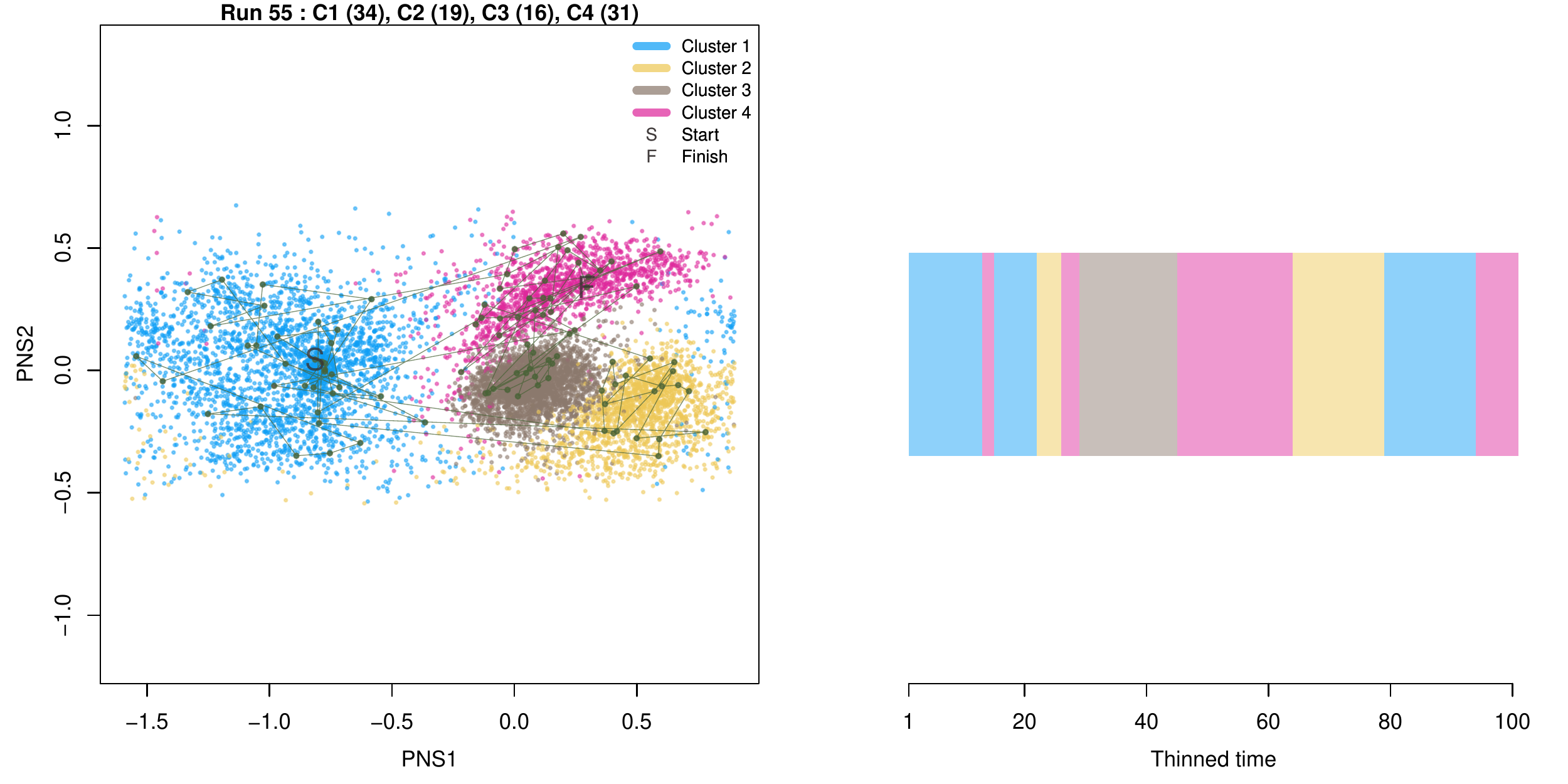}}
      \centering \subfigure[All runs]
   {\includegraphics[scale = 0.2]{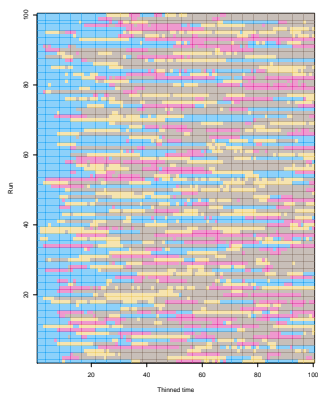}}
   \caption{Visit history.}
   \label{fig:visit_history1}
\end{figure}

A graphical view of shape changing pattern within these four groups over time can be seen as 
in Figure \ref{fig:visit_history1}(a) for run 55.
This run changes shape regularly
but some other runs have different patterns,
for example run 66 changes very much during the middle times,
run 25 does not visit the yellow cluster, and so on. All runs are shown in Figure \ref{fig:visit_history1}(b). 

Since all 100 runs behave with different dynamic patterns,
we ran cluster analysis on 100 transition matrices within these four locations.
To this end we estimate transition matrices for each run, $T_j, j = 1, \ldots, 100$, using empirical probabilities
and we use the Hellinger distance on transition matrices defined by
$$
H(T_1, T_2)
=
\frac{1}{\sqrt{2}}
\big\| \sqrt{T_1} - \sqrt{T_2} \big\|.
$$
The overall transition matrix $P$ estimated from all the runs is given in Table \ref{tab:trans.avg}, indicating the conditional 
probability from a cluster (rows) to another cluster (columns). The transition jumps are evaluated at each 100 time points on the original scale, i.e. for the thinned data. 
The overall equilibrium distribution of the chain is given in 
Table \ref{tab:equib}, obtained from $P^n$ as $n \to \infty$.  
Overall we see that more time is spent in Cluster 3 and less in Cluster 4. 
There is a preference to move from Cluster 1 to either Cluster 2 or 4 first, before then transitioning 
to Cluster 3 as the next most likely move. So, we can see that the movement between the states is asymmetric, 
and there are preferred orderings of transitions between states.  

\begin{table}[htbp]
   \caption{Overall average transition matrix.}
   \label{tab:trans.avg}
\begin{center}
\begin{tabular}{c|c|c|c|c}
\hline
 & 1 & 2 & 3 & 4 \\
\hline
1 & 0.8628 & 0.0712 & 0.0135 & 0.0525 \\
2 & 0.0744 & 0.7480 & 0.1608 & 0.0168 \\
3 & 0.0069 & 0.0893 & 0.8501 & 0.0537 \\
4 & 0.0655 & 0.0178 & 0.1588 & 0.7578 \\
\hline
\end{tabular}\\
\end{center}
\end{table}

\begin{figure}[htbp]
   \centering
   {\includegraphics[scale = 0.35]{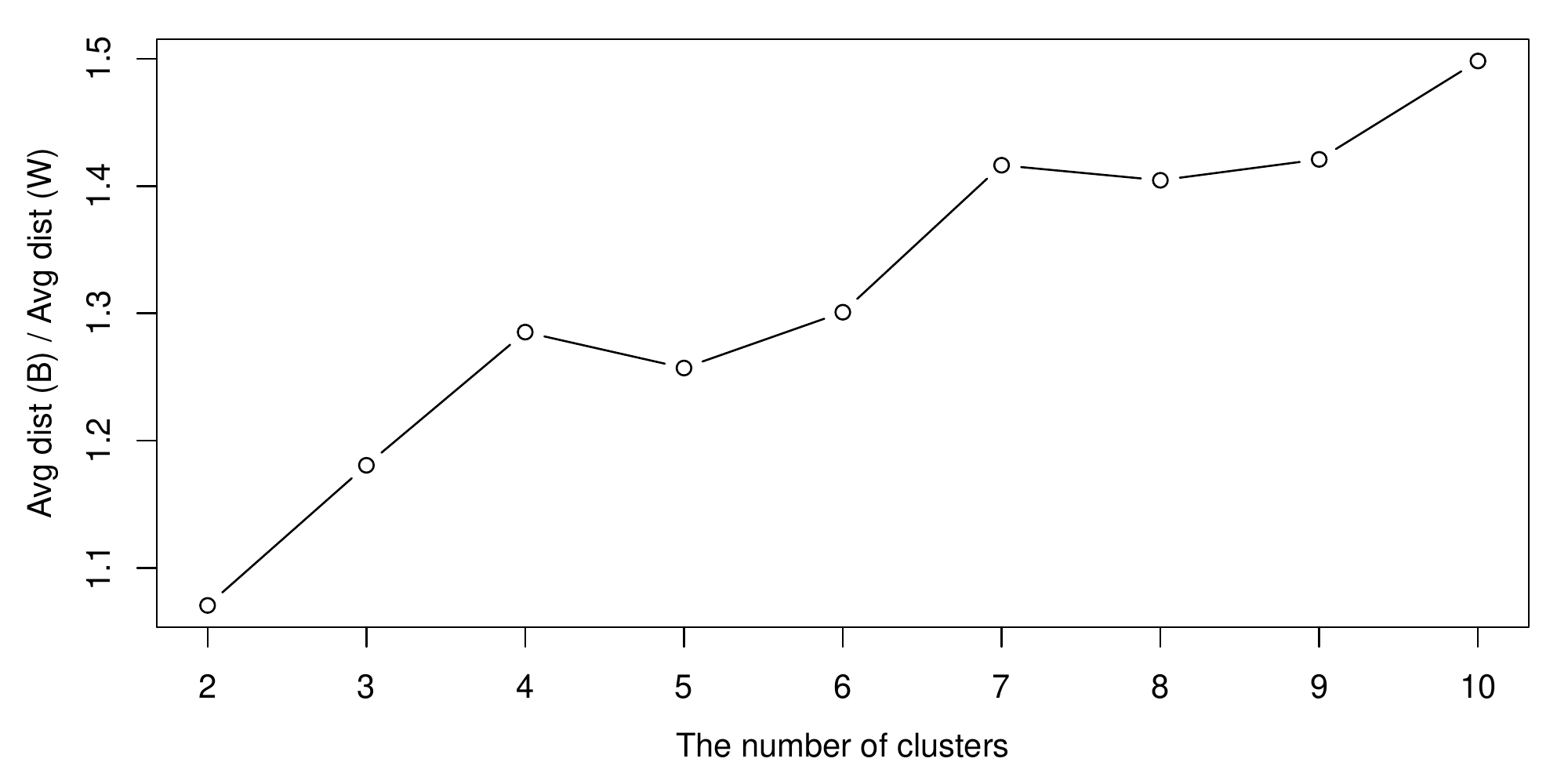}}
   \centering
   {\includegraphics[scale = 0.18]{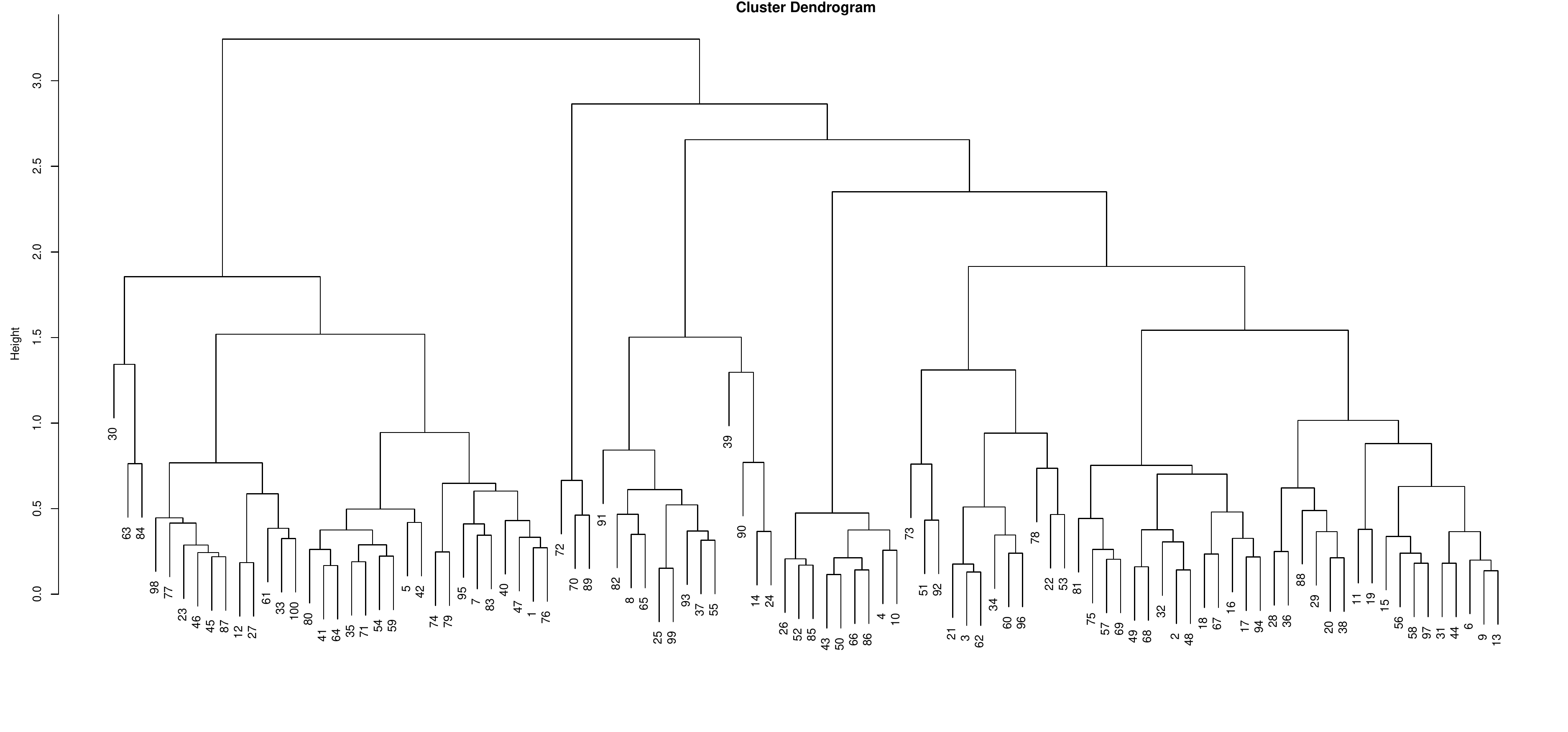}}
   \centering
   {\includegraphics[scale = 0.36]{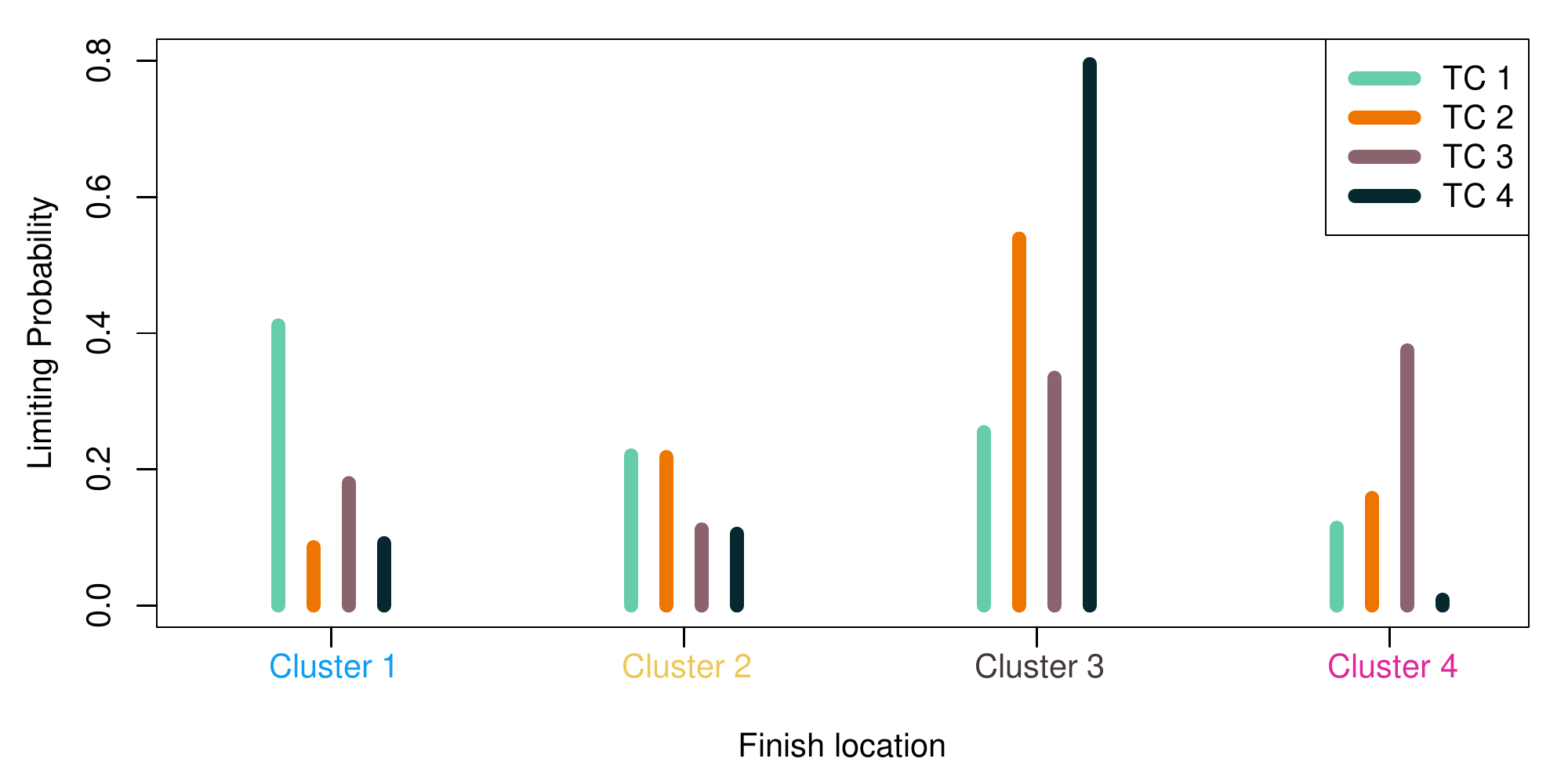}}
   \caption{TC1 - TC4 denote temporal cluster labels after clustering transition matrices
   and Cluster 1 - Cluster 4 on $x$-axis denote the finishing location labels after clustering on the $S^{10}$ data.}
   \label{fig:clust.trans}
\end{figure}

\begin{table}[htbp]
\begin{center}
\begin{tabular}{c|c|c|c|c}
\hline
 & 1 & 2 & 3 & 4 \\
\hline
Overall & 0.213 & 0.218 & 0.416 & 0.153\\
\hline
TC1 & 0.411 & 0.220 & 0.254 & 0.114\\
TC2 & 0.085 & 0.218 & 0.540 & 0.158\\ 
TC3 & 0.180 & 0.112 & 0.334 & 0.374\\
TC4 & 0.091 & 0.105 & 0.795 & 0.008\\
\hline
\end{tabular}
\end{center}
   \caption{Equilibrium probabilities for Cluster 1,2,3,4, for the overall dataset and for runs within each temporal cluster 
TC1,TC2,TC3,TC4.}
   \label{tab:equib}
\end{table}

According to the first panel in Figure \ref{fig:clust.trans}
showing the average distance between clusters divided by average distance within clusters,
we split the transition matrices into four groups (temporal clusters) TC1 to TC4, by hierarchical clustering using Ward's method
with the Hellinger distance.
Focusing on the finishing locations where runs terminate,
we estimated the final probability for each location.
In the bottom panel, the labels Cluster 1 to Cluster 4 on $x$-axis denote the finish location labels
which correspond to four groups in Figure \ref{fig:clust} (b).
Again we see that Cluster 3 is seen as finish location with high probability overall
and relatively fewer runs terminate at Cluster 4.

We also estimate the transition matrices with TC1,TC2,TC3 and TC4 and display the equilibrium probabilities in Table \ref{tab:equib}. 
Simulation runs in TC1 have a relatively strong pull to Cluster 1, 
runs in TC2 have a pull towards Cluster 3, 
runs in TC3 have a pull towards Clusters 3 and 4, and runs in 
TC4 have a very strong pull towards Cluster 3. This behaviour 
is also seen in the estimated probability of final locations in Figure \ref{fig:clust.trans}. 
 
Hence it is of interest that the molecular dynamics simulations do exhibit different behaviours in the runs, with TC1 being particularly different from the other temporal clusters.  

The analysis is dependent on the choice of temporal thinning. We have chosen to thin to every 100 observations due to the very fine temporal resolution in the original data. 
There would be some small periods spent in other clusters between the thinned values, but the choice of 100 gives a good compromise between excessive long times spent
in a cluster (too little thinning) versus brief/missed visits to clusters (too much thinning). 

\section{Discussion}
We have developed the method of principal nested shape spaces as an extension of principal nested spheres 
\citet{JDM12}, and the peptide application clearly demonstrates the utility of the method. The technique has been applied 
to other datasets, including molecular dynamics simulations of enzyme data. Again useful insights are obtained from the backwards 
method for the enzymes data that are different from using conventional tangent space shape PCA, which is a forwards method that requires estimation of the mean shape at the outset. 
The general approach to backwards PCA and principal nested manifold relations is discussed by \citet{Damomarr14}.  
Further related work includes \citet{Pennec17} who discusses barycentric sub-spaces, where sub-manifolds are generated from 
weighted means of reference points. 

The data are available at:
\begin{center}
{\tt http://www.maths.nottingham.ac.uk/$\sim$ild/pnss}
\end{center}

\bibliographystyle{apalike}
\bibliography{pep}

\vskip 2cm

\section*{Appendix 1}
\subsection*{Principal nested sub-shape-spaces for $m$-dimensional data}
Let $\tilde X$ in $\mathbb{R}^m$ denote $k > m$ labelled landmarks that are not all identical, and write $X$ for the pre-shape 
of the configuration obtained from $\tilde X$ by removing the effects of translation and scaling. The pre-shape $X$ is 
a $(k-1)\times m$ matrix of unit norm, and the pre-shape sphere is denoted $\mathcal{S}_m^k$.

The tangent space $\mathcal T_X(\mathcal{S}_m^k)$ to $\mathcal{S}_m^k$ at any $X\in\mathcal{S}_m^k$ is the subspace of $\mathcal{M}(k-1,m)$ given by
\[\mathcal T_X(\mathcal{S}_m^k)=\{Z\in\mathcal{M}(k-1,m)\mid\hbox{tr}(X^\top Z)=0\},\]
where $\mathcal{M}(k-1,m)$ denotes the space of all $(k-1)\times m$ matrices. In terms of the quotient map from the pre-shape sphere $\mathcal{S}_m^k$ to the Kendall shape space $\Sigma_m^k$ of configurations in $\mathbb{R}^m$ of $k$ labelled landmarks, we can decompose $\mathcal T_X(\mathcal{S}_m^k)$ into the direct sum of two orthogonal subspaces, one of which is tangent to the equivalence class of $X$. We denote this subspace of $\mathcal T_X(\mathcal{S}_m^k)$ by $\mathcal{V}_X$. Its orthogonal complement $\mathcal{H}_X$ is the Procrustean, or the horizontal, sub-tangent space at $X$. It is known  that \citep{Kentmard01}
\begin{eqnarray*}
\mathcal{V}_X&=&\{XA\mid A\hbox{ is } m\times m\hbox{ skew-symmetric matrix}\}\\
\mathcal{H}_X&=&\{Z\in\mathcal T_X(\mathcal{S}_m^k)\mid X^\top Z\hbox{ is symmetric}\},
\end{eqnarray*}
and that $\mathcal{H}_X$ is isometric with the tangent space at $[X]$ to the shape space $\Sigma_m^k$.

\vskip 6pt
For any $1\leqslant j_1<j_2\leqslant m$, write $E_{j_1j_2}$ for the $m\times m$ skew-symmetric matrix with $(j_1,j_2)$ component 1, $(j_2,j_1)$ component $-1$ and 0 otherwise. Then, for any pre-shape $X_0\in\mathcal{S}_m^k$ with rank$(X_0)=m$, 
\[\hbox{tr}\left((X_0E_{j_1,j_2})^\top(X_0E_{j_1,j_2})\right)=\|X_0E_{j_1,j_2}\|^2>0,\]
so that $X_{0;j_1,j_2}=X_0E_{j_1,j_2}/\|X_0E_{j_1,j_2}\|\in\mathcal{S}_m^k$. Moreover, 
$\{X_{0;j_1,j_2}\mid1\leqslant j_1<j_2\leqslant m\}$ spans 
$\mathcal{V}_{X_0}$. 

The subset of $\mathcal{S}_m^k$ orthogonal to the Euclidean subspace spanned by $\{X_{0;j_1,j_2}\mid1\leqslant j_1<j_2\leqslant m\}$ is a great sphere of dimension $m\times(k-1)-m(m-1)/2-1$. 
We denote this subsphere by $\mathcal{S}_{X_0}$, that is,
\[\mathcal{S}_{X_0}=\{S\in\mathcal{S}_m^k\mid \rho(S,X_{0;j_1,j_2})=\pi/2, \forall1\leqslant j_1<j_2\leqslant m\}.\]
Clearly, $X_0\in\mathcal{S}_{X_0}$ and the dimension of $\mathcal{S}_{X_0}$ is the same as that of the shape space $\Sigma_m^k$. It can be checked that $\mathcal{S}_{X_0}$ is the image of $\mathcal{H}_{X_0}$ under the exponential map at $X_0$.  

For any pre-shape $X\in\mathcal{S}_m^k\setminus\{X_0\}$, we apply the ordinary Procrustes analysis procedure to choose $S_X=XR_X\in[X]$, where 
\[R_X=\argmin_{R\in SO(m)}\rho(X_0,XR).\]
\begin{enumerate}
\item[$\bullet$] The resulting $S_X$ is usually called the Procrustes fit to $X_0$ of the original configuration; 
\item[$\bullet$] $S_X$ has the same shape as $X$; 
\item[$\bullet$] $X_0^\top S_X$ is symmetric; 
\item[$\bullet$] $\rho(X_0,S_X)\leqslant\pi/2$; and 
\item[$\bullet$] the spherical distance distance, $\rho(X_0,S_X)$, between $S_X$ and $X_0$ is the same as the induced Riemannian shape distance between the shapes of $X_0$ and $X$. 
\end{enumerate}

The fact that $X_0^\top S_X$ is a symmetric matrix implies that, for any $1\leqslant j_1<j_2\leqslant m$,
\[\langle X_{0;j_1,j_2},S_X\rangle=\hbox{tr}((X_0E_{j_1,j_2})^\top S_X)=-\hbox{tr}(E_{j_1,j_2}X_0^\top S_X)=0,\]
so that $\rho(X_{0;j_1,j_2},S_X)=\pi/2$. It follows that $S_X\in\mathcal{S}_{X_0}$ and, in particular, that the intersection $\{XR\mid R\in SO(m)\}\bigcap\mathcal{S}_{X_0}$ is always non-empty. Since $\rho(X_0,S_X)\leqslant\pi/2$, all such $S_X$ lie in the same half sphere of $\mathcal{S}_{X_0}$. 

For given $X_0$ and $X$, the corresponding $R_X$, and hence the corresponding $S_X$, is not necessarily unique. The uniqueness of $S_X$, or $R_X$, is equivalent to the shape $[X]$ being within the non-singular part of $\Sigma_m^k$ but outside the cut locus of the shape $[X_0]$ \citep{Le91}. Hence, the subset
\[\mathcal{B}_{X_0}=\{S_X\in\mathcal{S}_{X_0}\mid X\in \mathcal{S}_m^k\hbox{ and }S_X\hbox{ is unique}\}\] 
of $\mathcal{S}_{X_0}$ is topologically a ball in $\mathcal{S}_{X_0}$ containing $X_0$ and is bijective, under the quotient map, with the subset
\[\mathcal{B}_{[X_0]}=\{[X]\in\Sigma_m^k\mid[X]\hbox{ is non-singular and not in the cut locus of }[X_0]\}\]
of $\Sigma_m^k$. Note that $[X]$ being non-singular is equivalent to it being shape of a configuration whose pre-shape matrix $X$ has rank at least $m-1$. Since the quotient map is differentiable and since $\mathcal{B}_{X_0}$ and $\mathcal{B}_{[X_0]}$ are respectively submanifolds of $\mathcal{S}_m^k$ and $\Sigma_m^k$, it follows that these two sets are diffeomorphic with each other. Thus, any sub-manifold in $\mathcal{B}_{X_0}\subset\mathcal{S}_{X_0}$ is mapped to a sub-manifold in $\mathcal{B}_{[X_0]}\subset\Sigma_m^k$. In particular, the intersection of $\mathcal{B}_{X_0}$ with a subsphere of $\mathcal{S}_{X_0}$ is mapped to a sub-manifold of $\Sigma_m^k$, so that the intersection of $\mathcal{B}_{X_0}$ with a sequence of nested subspheres of $\mathcal{S}_{X_0}$ is mapped to a sequence of nested sub-manifolds of $\Sigma_m^k$. However that these maps are not isometric. Nevertheless, the great circle arc through $X_0$ is mapped to a geodesic in $\Sigma_m^k$ through $[X_0]$.

\section*{Appendix 2}
\subsection*{Approximate PNSS using PCA}
Consider configurations $\tilde{X}_i$, $i=1,\ldots,n$ in $\mathbb{R}^m$ of $k$ labelled landmarks with Procrustes mean shape $\bar X$. Their pre-shape matrices are $X_i$, $i=1,\ldots,n$, and their Procrustes fits to $\bar X$ are $S_i=S_{X_i}$, $i=1,\ldots,n$. Let $T_i$, $i=1,\ldots,n$, be the Procrustes tangent projection of $X_i$ at $\bar X$ 
\citep[(4.33)]{Drydmard16}, i.e.
\[T_i=S_i-\langle\bar X,S_i\rangle\,\bar X=S_i-\hbox{tr}(\bar X^\top S_i)\,\bar X\in\mathcal{H}_{\bar X}\subset \mathcal T_{\bar X}(\mathcal{S}_m^k),\]
and  $v_j$ be the unit eigenvector corresponding to the $j$th largest eigenvalue of the covariance matrix of $T_i$, $i=1,\ldots,n$,
\[\mbox{cov} =\frac{1}{n} \sum_{i = 1}^n{\rm vec}(T_i- \bar T){\rm vec}( T_i - \bar T)^\top, \]
where $y = {\rm vec}(Y)$ is the vectorize operation of stacking the columns $Y$ into a single vector $y$, 
$\bar T = \frac{1}{n} \sum_{i=1}^n T_i$ and $V_j = {\rm vec}_m^{-1}(v_j)$, where ${\rm vec}_m^{-1}(y) = Y$  is the inverse vectorise operator forming a 
matrix $Y$ of $m$ columns. 
Then, $\langle\bar X, V_j\rangle=0$ and $V_j\in\mathcal{H}_{\bar X}$. For any $1\leqslant p< m(k-1)-m(m-1)/2-1$, the $p$-dimensional unit sphere $\mathcal{S}_{\bar X;V_1,\ldots,V_p}$ in the $(p + 1)$-dimensional linear subspace spanned by $\bar X,V_1,\ldots,V_p$ is a $p$-dimensional subsphere of $\mathcal{S}_{\bar X}$. A minimal practical requirement is that the first $p$ eigenvalues will be strictly positive. If we choose $p$ such that the first $p$ principal components explain a high proportion of the total variation of the data then, without much loss of information, the use of the projection of $S_i$ to $\mathcal{S}_{\bar X;V_1,\ldots,V_p}$ as the input for the principal nested sub-shape-spaces analysis will reduce the initial dimension of the sphere and improve the speed of computation.

The projection of $S_i$ to $\mathcal{S}_{\bar X;V_1,\ldots,V_p}$ can be approximated by using the shape principal component scores, where the shape principal component score for the $i$th configuration on the $j$th principal component is given by 
$\tilde\lambda_{ij}=\langle T_i,V_j\rangle, i=1,\ldots,n, j=1,\ldots,m(k-1)-m(m-1)/2-1$.
For this, let
\[W_i=\rho(\bar X,S_i)\frac{T_i}{\|T_i\|}\in\mathcal T_{\bar X}(\mathcal{S}_m^k).\]
Then, $W_i$ is the image of $S_i$ under the inverse exponential map of $\mathcal{S}_m^k$ at $\bar X$, it lies in $\mathcal{H}_{\bar X}$ and, in particular, $\|W_i\|=\rho(\bar X,S_i)$. By writing
\[\lambda_{ij}=\langle W_i,V_j\rangle=\frac{\rho(\bar X,S_i)}{\|T_i\|}\tilde\lambda_{ij},\]
the projection of $W_i$ in the subspace, spanned by $V_1,\ldots,V_p$, of the tangent space $\mathcal T_{\bar X}(\mathcal{S}_m^k)$ is
\begin{equation}\label{eq:tancoords}
U_i=\sum_{j=1}^p\lambda_{ij}V_j.
\end{equation}
The image $S^*_i$ of $U_i$ under the exponential map back in the pre-shape sphere is
\begin{eqnarray*}
S_i^*&=&\bar X\cos(\|U_i\|)+\frac{U_i}{\|U_i\|}\sin(\|U_i\|)\\
&=&\cos(\|U_i\|)\,\bar X+\frac{\sin(\|U_i\|)}{\|U_i\|}\sum_{j=1}^p\lambda_{ij}V_j\in\mathcal{S}_{\bar X; V_1,\ldots,V_p}
\end{eqnarray*}
where, if $\|U_i\|=0$, we take $S^*_i=(1,0,\ldots,0)$. When $S_i$ is sufficiently close to $\mathcal{S}_{\bar X;V_1,\ldots,V_p}$, $S^*_i$ gives a good approximation to the projection of $S_i$ to $\mathcal{S}_{\bar X;V_1,\ldots,V_p}$. 

Since $\bar X, V_1,\cdots,V_p$ are orthonormal, they form a basis of the $(p + 1)$-dimensional linear subspace spanned by themselves and, in terms of this new basis, $S^*_i$ can be expressed as
\[\left(\cos(\|U_i\|),\frac{\sin(\|U_i\|)}{\|U_i\|}\lambda_{i1},\cdots,\frac{\sin(\|U_i\|)}{\|U_i\|}\lambda_{ip}\right).\]  
This representation will simplify computation. Conversely, for a given $G = (G_1, G_2, \ldots, G_{p + 1})^\top \in\mathcal{S}_{\bar X;V_1,\ldots,V_p}$, its principal component scores are
\begin{equation}\label{eq:pc.rev}
\frac{s}{\sin(s)}\left(G_2,G_3,\ldots,G_{p + 1}\right)^\top
\end{equation}
where $s = \cos^{-1}(G_1)$. 

\vskip 2cm 
$\;$
\end{document}